\documentclass[manuscript]{aastex}

\usepackage{amsmath}
\usepackage{graphicx}
\usepackage{natbib}
\usepackage{rotating}
\usepackage{leqno}
\usepackage{longtable}
\usepackage{txfonts}

\slugcomment{}

\shorttitle{Study of Metal-Poor Binaries}
\shortauthors{Benamati et al.}

\begin{document}

\title{A Combined Astrometric and Spectroscopic Study of Metal-Poor Binaries}

\author{L. Benamati\altaffilmark{1,2}, A. Sozzetti\altaffilmark{3}, N.C. Santos\altaffilmark{1,2} }

\and

\author{D.W. Latham\altaffilmark{4}}

\email{Lisa.Benamati@astro.up.pt}

\altaffiltext{1}{Centro de Astrofisica, Universidade do Porto, Rua das Estrelas, 4150-762 Porto, Portugal}
\altaffiltext{2}{Departamento de F\'{\i}sica e Astronomia, Faculdade de Ci\^encias, Universidade do Porto, Portugal}
\altaffiltext{3}{INAF- Osservatorio Astrofisico di Torino, 10025 Pino Torinese, Italy}
\altaffiltext{4}{Harvard-Smithsonian Center for Astrophysics, 60 Garden Street, Cambridge, Massachusetts 02138, USA}

\begin{abstract}
In this work we present a study of binary systems in a metal-poor sample of solar type stars. 
The stars analyzed were rejected from two planet search samples because they were found to be binaries. 
Using available radial velocity and Hipparcos astrometric data, we apply different methods to find, for every binary system, 
a possible range of solutions for the mass of the companion and its orbital period. In one case we find that the solution depends 
on the Hipparcos data used: the old and new reductions give different results. 
Some candidate low-mass companions are found, including some close to the brown dwarf regime.
\end{abstract}


\section{Introduction}
Since the discovery of the first extrasolar planet in 1995 (51 Peg b, \citealt{may95}) 
the search for the extrasolar planetary systems has witnessed spectacular successes. Today the number of discovered planets is around 900, 
but, even if the number is relatively high, theories of planet formation and evolution is still under discussion. Therefore, the 
study of the frequency of different types of planets around stars with different properties (metallicity, mass, etc.) can 
help us understanding better the processes of planet formation and evolution. \\ 
In particular, the evidence for a strong direct dependency of giant planet frequency with stellar metallicity [Fe/H] (e.g., \citealt{san04}; \citealt{fis05})
has motivated the design of Doppler surveys focus on the search for planets orbiting metal-deficient stars (\citealt{san11,soz09}), 
usually not observed in large numbers as part of the most successful decade-long RV survey programs. In these samples some stars were found 
to be unsuitable targets for a planet search for different reasons. 
They can, however, be useful for a number of other studies. In particular, the discovery of several binaries in these programs allow us to address 
the study of metal-poor binary stars. \\
The study of differences between binary frequencies for metal-poor and metal rich stars has a long history. For example, \citet{lat02} 
found that the period distribution does not present a strong correlation with [Fe/H]. This suggests that the metallicity 
has little influence over the fragmentation process that leads to the formation of short-period binaries. On the other hand, \citet{abt09} 
asserted that the absolute fractions of short-period binaries (P $<$ 100 days) is much smaller for the metal-poor stars than for the metal-rich stars. 
Moreover they concluded that the more probable period for the metal-poor stars is around 875 days, while for the metal-rich stars it 
is around 25 days. Therefore, the results are in disagreement and show that the impact of $[Fe/H]$ in the formation of binary stars is still a topic worthy investigatin. \\ 
Another study, focused on the mass ratio distributions\footnote{The mass ratio is denoted by $q = \frac{M_{2}}{M_{1}}$, where $M_{2}$ is 
the mass of the secondary star and $M_{1}$ the mass of the primary}, affirms that this ratio (for solar-type binaries) is approximately 
flat and uniform, even if the distribution shows a peak around 0.2 (\citealt{duq91}). This study includes the VLMC (very low mass companion) 
binaries (with $q < 0.1$), too. \citet{duq91} find that there are $57 \%$ of binary systems with a mass ratio higher than 0.1 and $43 \%$ of 
apparently single stars. Of these single stars, $(8 \pm 6) \%$ most probably have a VLMC. More recent analyses \citep{jan12,bat12} 
confirm this smooth distribution in the case of FGK stars. The way this trend depends on $[Fe/H]$ has, however, never been explored in detail. 
This is due to the relatively small numbers of known binaries among metal-poor stars. \\
In this paper we present a list of spectroscopic binary stars discovered in the context of two radial-velocity (RV) planet search programs (\citealt{san11,soz09}) 
(Sect.\ref{sec:sample}) focusing on metal-deficient F-G-K dwarfs. Using different methods (described in Section \ref{sec:methods}) 
we try to find a range of solutions for the period and the mass of the companion based on the available radial velocities and the combination of Hipparcos and Tycho astrometry. 
In Sect. \ref{sec:results} we show the results on a case by case basis. We conclude in Sect. \ref{sec:discus}.  

\section{The sample and data}
\label{sec:sample}

\begin{figure*}[p]
  \centering
  
  \begin{tabular}{c c c c}
    \includegraphics[width=7cm]{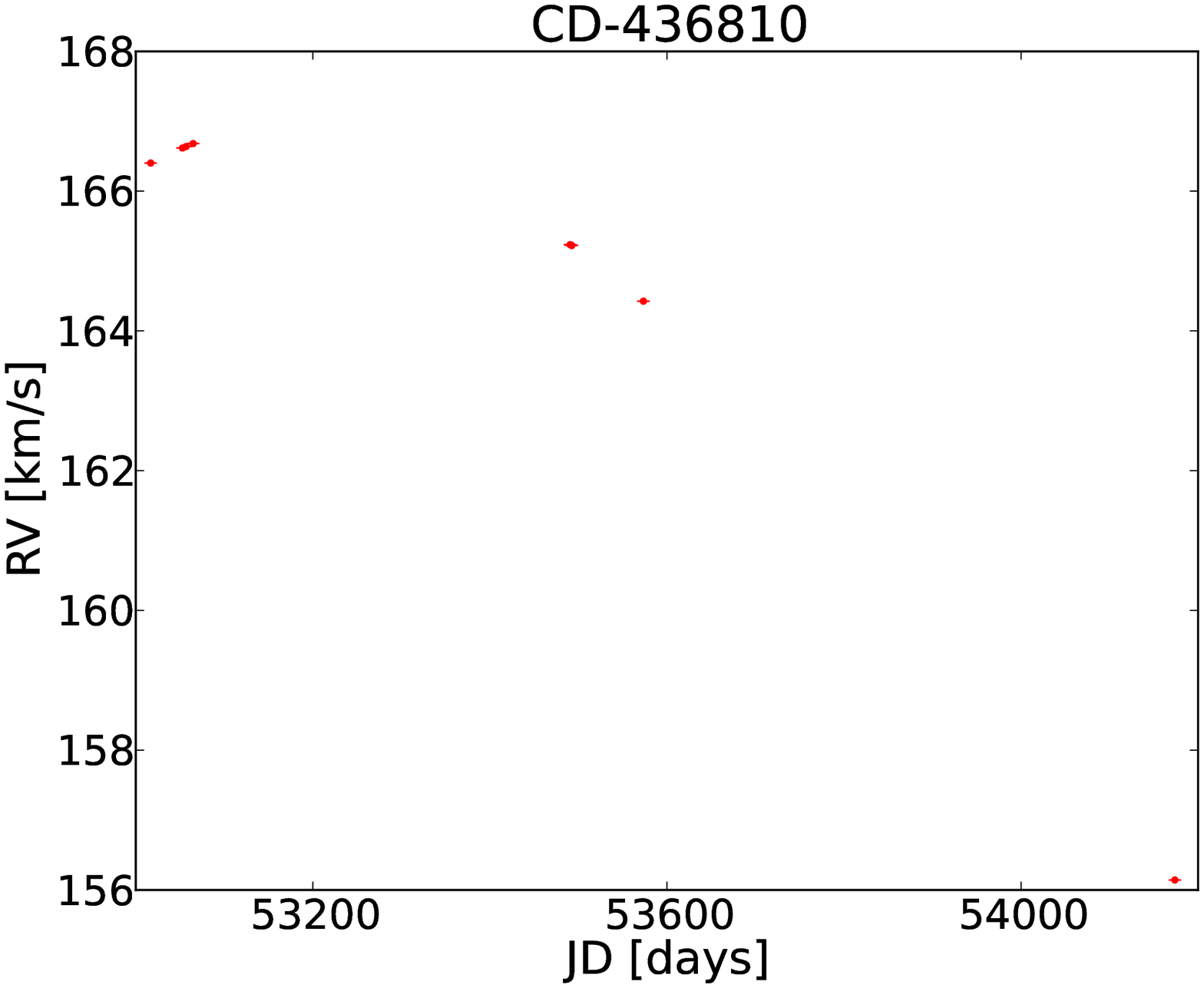}&

    \includegraphics[width=7cm]{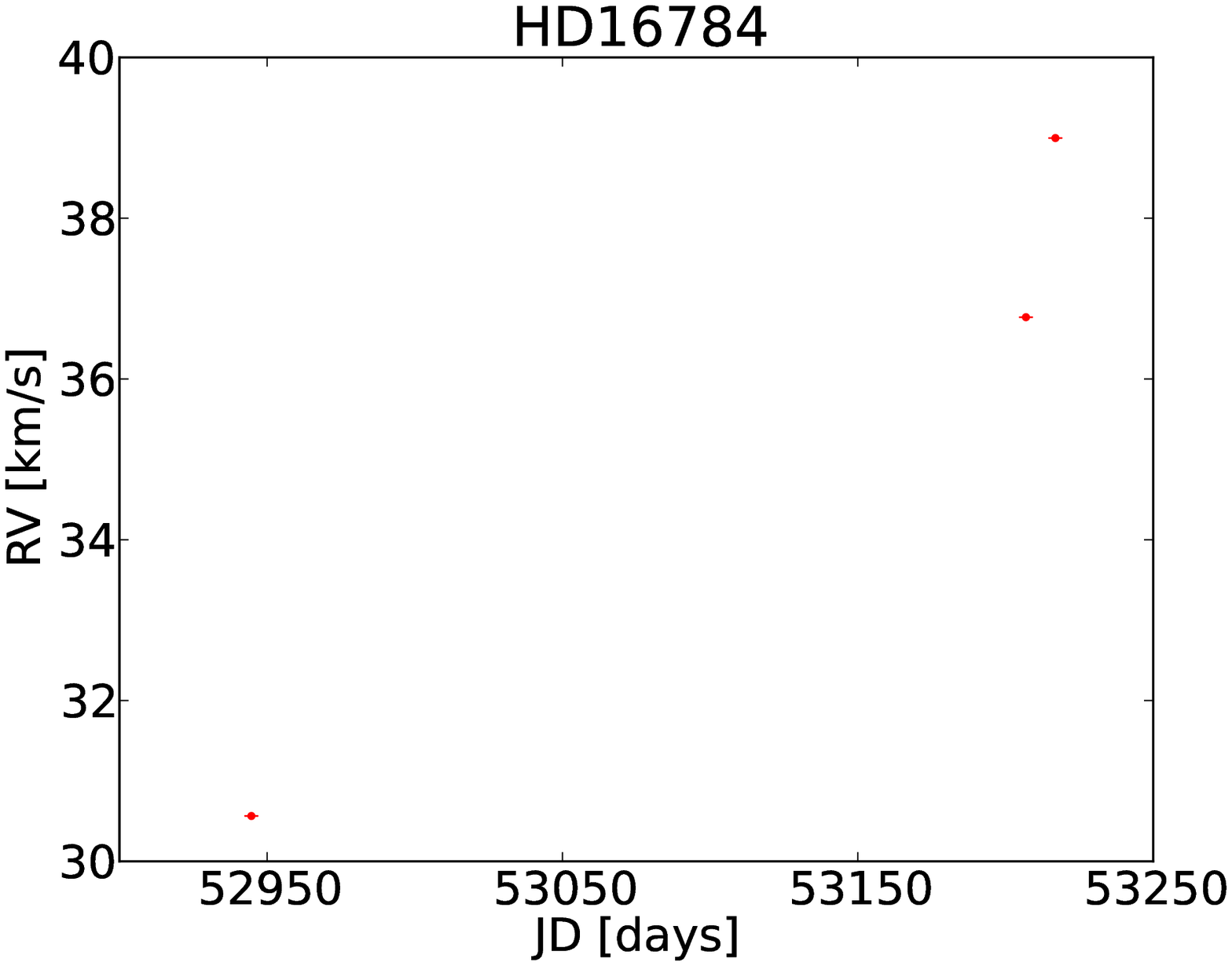}\\
    
    \includegraphics[width=7cm]{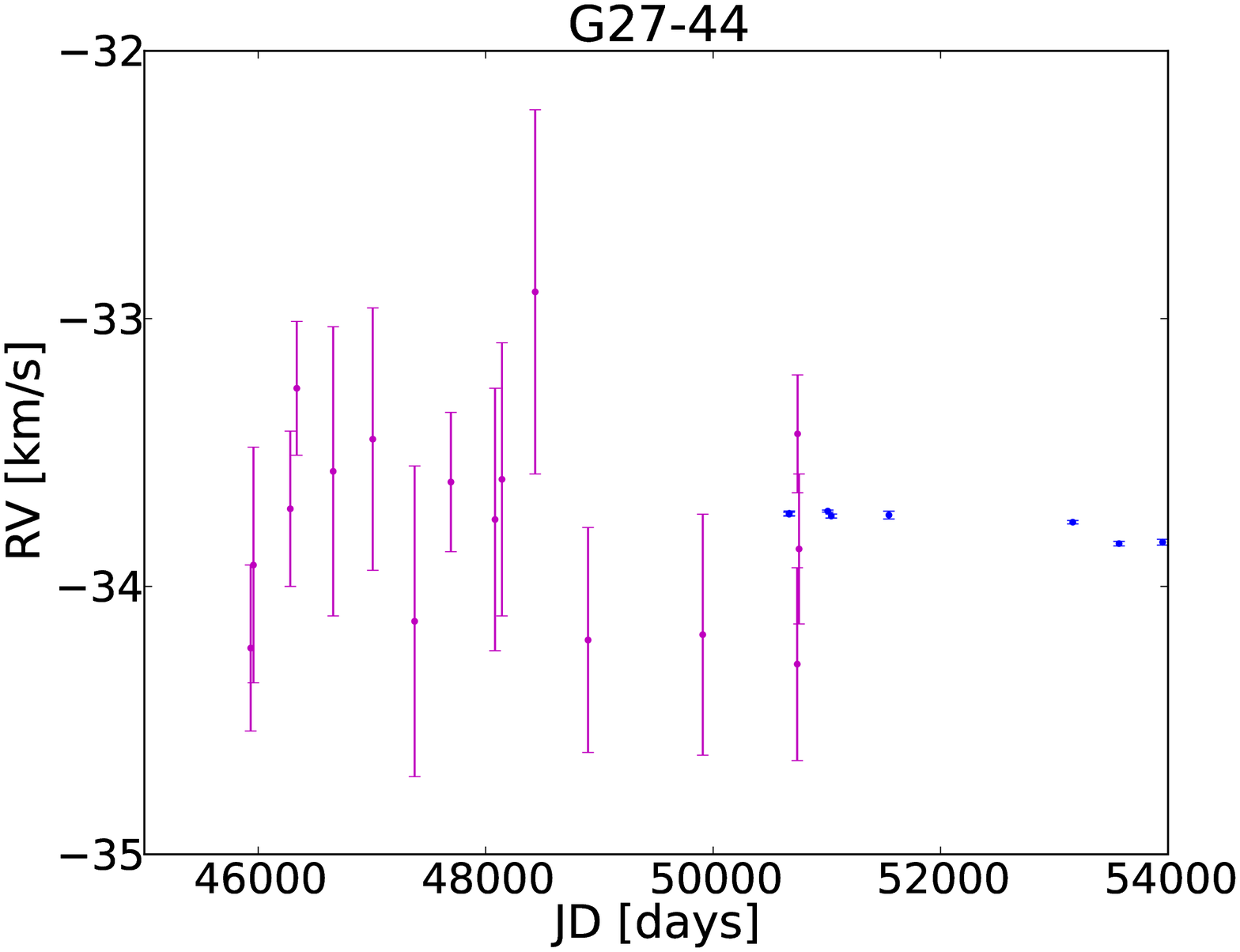}&

    \includegraphics[width=7cm]{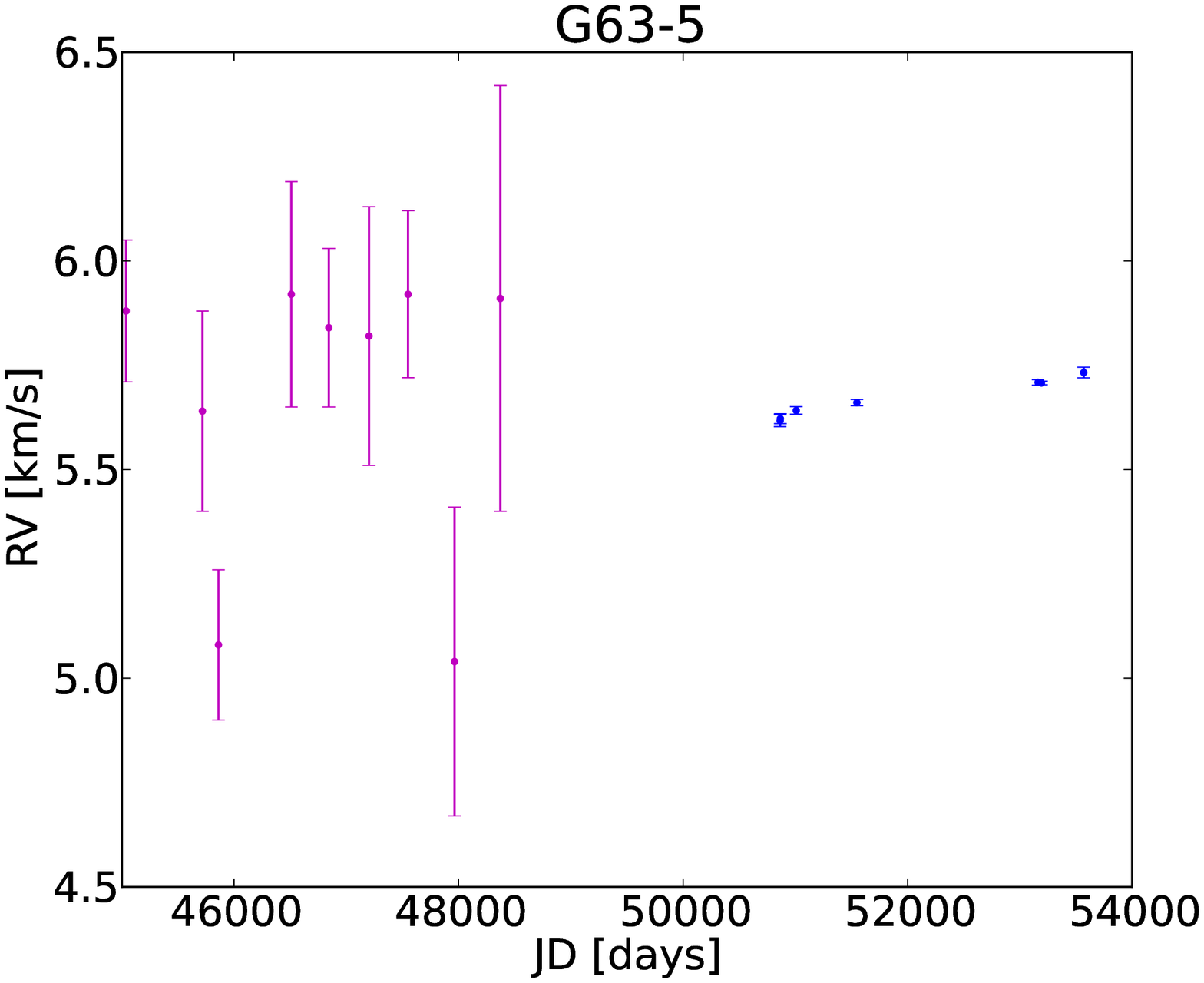}\\

    \includegraphics[width=7cm]{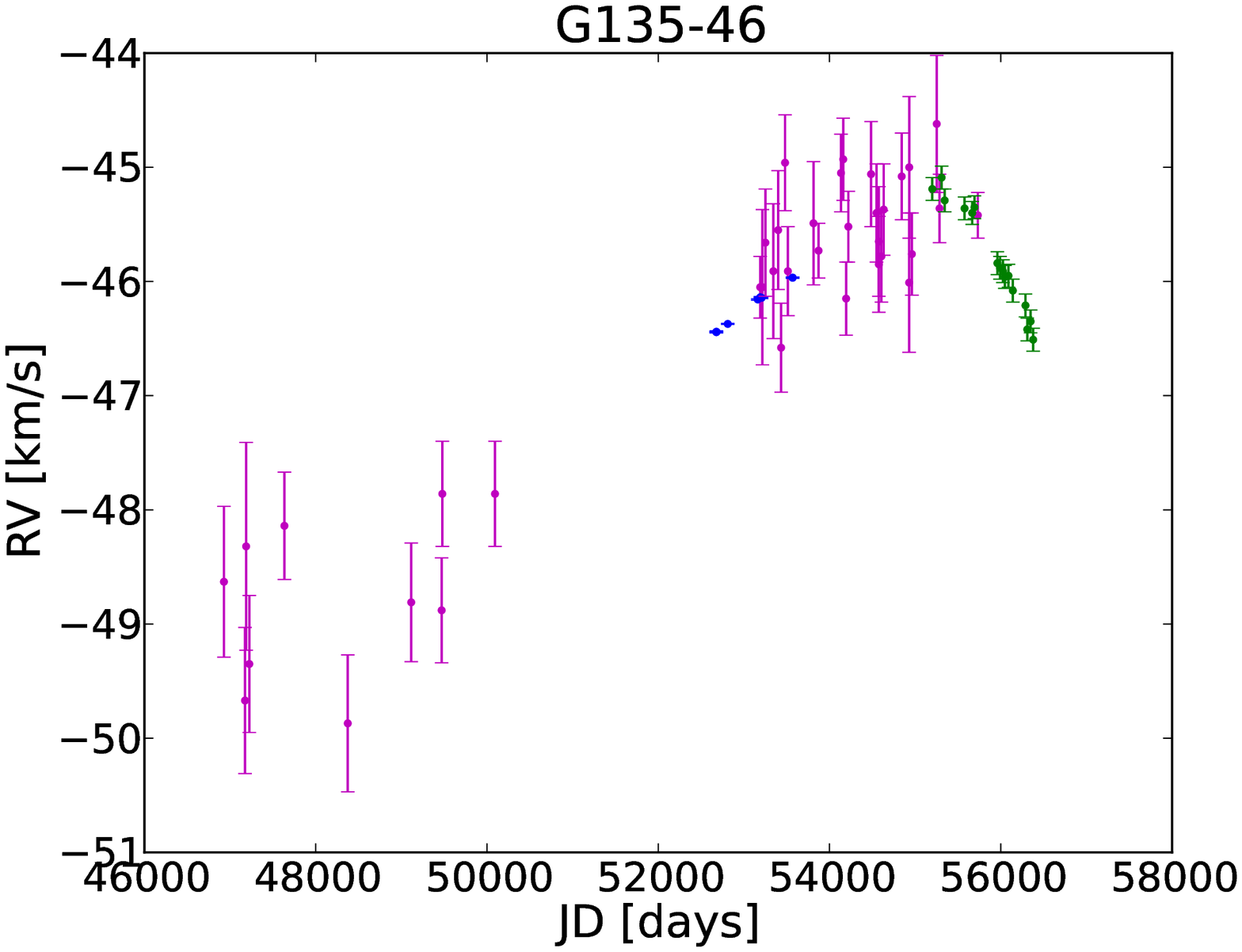}&

    \includegraphics[width=7cm]{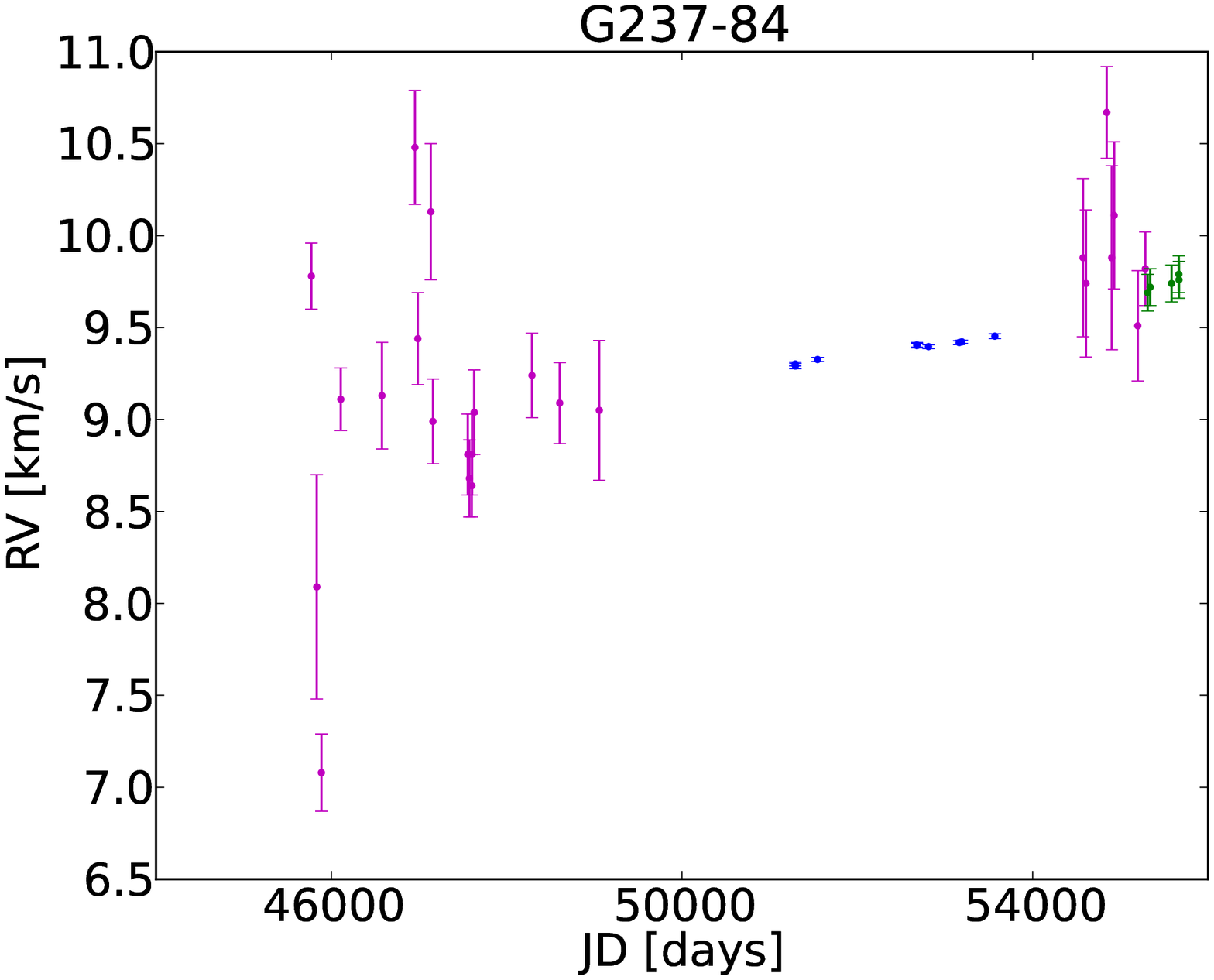}\\

    \includegraphics[width=7cm]{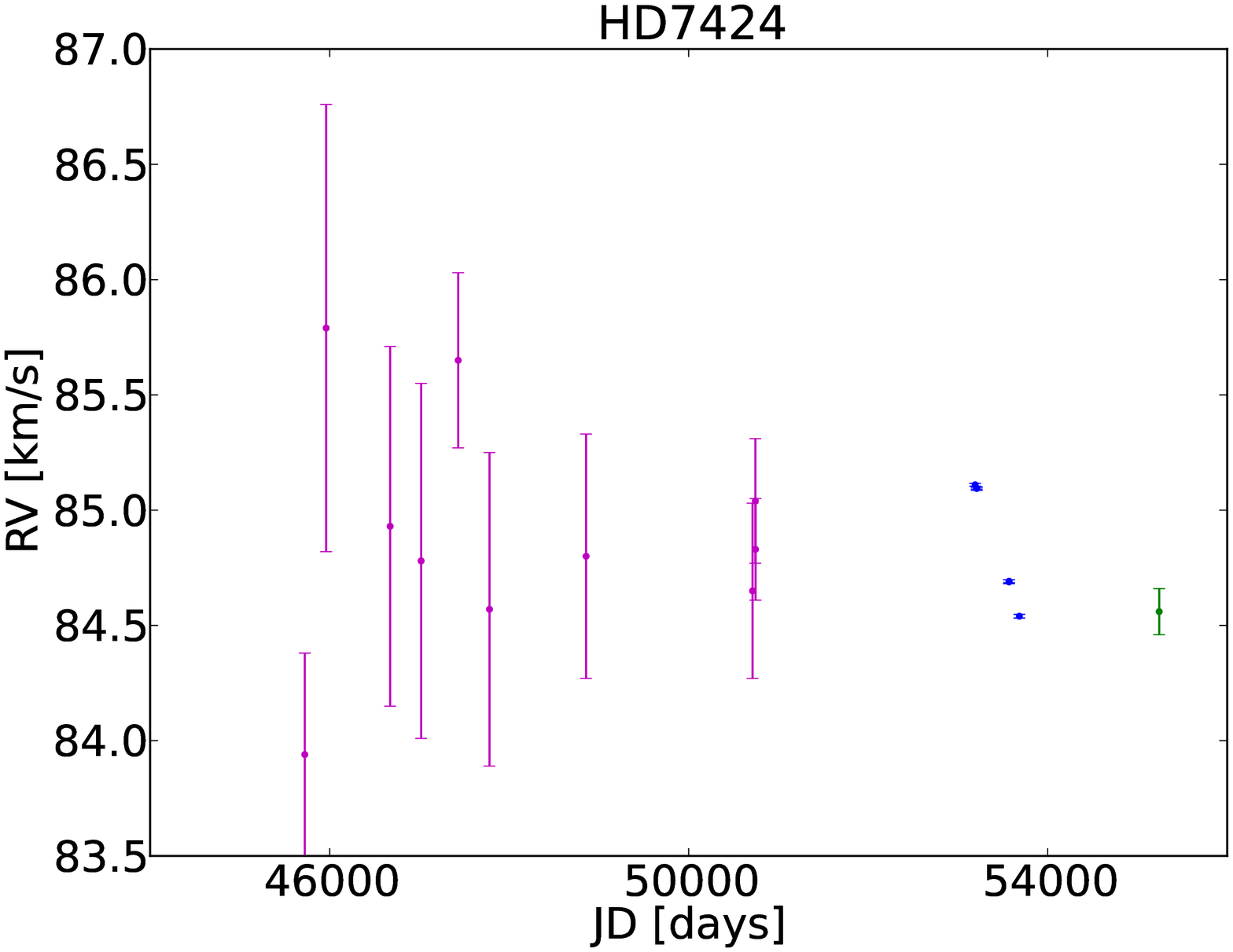}&

    \includegraphics[width=7cm]{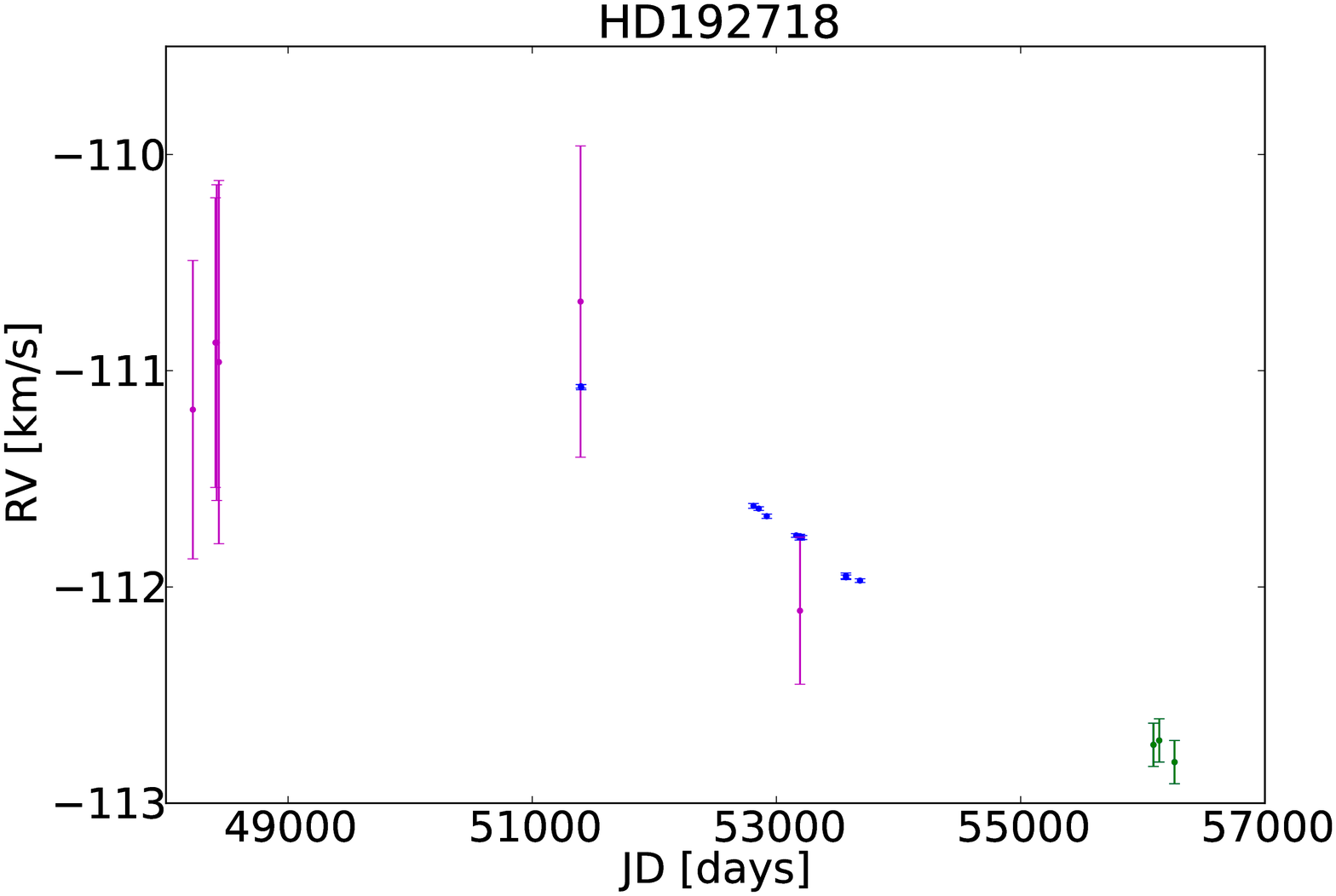}\\

    \\

  \end{tabular}
  
  \caption{RV measurements for the metal-poor binaries: red points indicate HARPS measurements, blue points HIRES data, magenta points the CfA DS measurements and green points the TRES velocities.}\label{fig:plotone}
\end{figure*}

Our sample of stars originally belongs to the Doppler surveys described in \citet{san11} and \citet{soz09}. These objects were successively 
removed once they were discovered to be SB1 spectroscopic binaries. No convincing orbital solution was found with the available RV data for the most part of the stars (except CD$-$436810 and G135$-$46). 
The list of stars analyzed in this paper is presented in Table \ref{table:1}. The detailed set of properties for this sample can be found in the above mentioned papers.\\
The RV measurements were obtained with the HARPS Spectrograph at La Silla (\citealt{may03}) from \citet{san11} and the HIRES Spectrograph on the Keck 1 
telescope at Mauna Kea in Hawaii (\citealt{vog94}) from \citet{soz09}. The HARPS data for CD$-$436810 and HD16784, spanning approximately three years and one year respectively, 
were gathered with a 2$-$3 m s$^{-1}$ precision. A full description of the data and observing strategy is provided in \citet{san11}. 
The RV measurements for these stars were obtained covering a timespan of about three years and less of one year. For the remaining stars, the RV data were obtained with HIRES 
(see \citet{soz09} for details) over a timespan of about three years (2003-2006). The precision of this data is typically $5-10$ m s$^{-1}$. 
Additional lower-precision RV time-series for all binaries (except CD$-$436810 and HD16784) were gathered with the CfA Digital Speedometers (\citealt{lat92})
and with TRES Echelle Spectrograph at the 1.5 meter Tillinghast telescope on Mt. Hopkins in Arizona (\citealt{fu08}). 
The typical TRES velocities precision is on the order of 100 m s$^{-1}$, while that of the CfA DS is $\sim0.5$ km s$^{-1}$ (see Table 2). 
These observations have a duration of anywhere between $\sim10$ years and more than 27 years.\\ 
To account for differences in the RV zero-points between datasets the full set of RV time-series is presented in the various panels of Fig.~\ref{fig:plotone}; in which the relative HIRES measurements for each star were shifted by the mean of the RV data from the CfA DS and TRES in order to bring them close to the common CfA DS + TRES system and small residual velocity offsets between the three systems were determined as free parameters in the best-fit orbital solution presented in table \ref{table:4}.
All HIRES data is public and published on the \citet{soz09} paper. The HARPS, CfA DS and TRES data are presented in Table \ref{table:m}. \\

\section{Methods}
\label{sec:methods}

Visual inspection of Fig. 1 clearly highlights the fact that for most systems in our sample the high-precision RV data at hand allow only to 
establish the existence of long-term (mostly linear) trends, and the lower-precision Doppler measurements (when available) usually cannot 
help to improve significantly the situation if one tries to combine them in order to derive full-fledged spectrocopic orbital solutions. 
However, it is still possible to glean some insight on the range of companion masses and periods for these stars by using, in addition to 
standard Keplerian orbit fitting algorithms when feasible, other ``non-standard'' statistical methods taking advantage of not only the 
RV data but also, when available, Hipparcos and Tycho-2 astrometry:

\begin{itemize}

\item The method described by \citet{mak05} was used to calculate the differences between the measured proper motion of Hipparcos and Tycho-2 ($\Delta\mu$), 
in addition to values for the acceleration in the proper motion ($\dot{\mu}$), when measured and reported in the Hipparcos catalogue. 
From these we used the statistical approach discussed by these authors to estimate the physical parameters of the systems, 
in particular the period and the mass of the secondary star.

\item When only a simple acceleration was detected in the Doppler studies (i.e. a linear RV trend), an estimate for the mass of the companion and the orbital period 
can also be derived using a statistical approach applying the method described by \citealt{tor99}. 

\end{itemize}

By combining the usually relatively loose constraints from each individual method, it is then possible to obtain tighter limits on the ranges of period 
and mass for the companions that are compatible with the available data. More details on the methodology used are provided below.

\subsection{Radial velocities}
\label{sec:RV}

\subsubsection{Orbital solution: fitting keplerian functions}
\label{sec:orbital}

The first step was to fit the radial velocity data with a Keplerian function. Only for two of the systems (G135$-$46 and CD$-$436810) it was possible to find a satisfactory Keplerian fit (see Sect. 4). 

The Keplerian function used for the measured radial velocity of the primary star is (\citealt{Hil01}):

\begin{equation} 
\label{eqn:e0} 
vr =\gamma + K[e*\cos{\omega} + \cos({\omega + \nu})] \\
\end{equation}   

where $\omega$ is the longitude of the periastron of the companion, $\nu$ is the true anomaly (both in degrees), $e$ is the eccentricity, 
$K$ the radial velocity semi-amplitude and $\gamma$ is the velocity of the center of mass of the system (both in m s$^{-1}$). 
Then, we calculated (when possible) the minimum mass of the companion using the Eq.\eqref{eqn:e01}:

\begin{equation} 
\label{eqn:e01} 
M_{min}(M_{J}) = \frac{1}{203.255} * M_{1}^{2/3}(M_{\sun}) * P^{1/3} (d) *K (m/s) * \sqrt{1-e^{2}}\\
\end{equation}   

\subsubsection{$d(RV)/dt$ method}
\label{sec:slope}
As mentioned above, another method we adopted uses the acceleration in the RV curve to estimate, from a statistical point of view, the mass of the companion. 
The method is fully described in \citet{tor99}, and was used for all cases for which the detected RV variation is in a good approximation a linear trend. 
This corresponds to situation where the period is very long and only a linear acceleration (a slope), $d(RV)/dt$, can be determined. 
However, for completeness, we have also decided to apply this method to the cases of G135$-$46 and in part for CD$-$436810, but only considering the data in the linear part of the curve. 
In these cases, we evaluate the expression for the secondary mass in terms of the measured RV acceleration, the distance estimate (D) and the angular separation ($\rho$):

\begin{equation} 
\label{eqn:e1} 
M_{2} = 5.341*10^{-6} (D\rho)^{2} \arrowvert \frac{d(RV)}{dt} \arrowvert \Phi(i,e,\omega,\varphi)  \\
\end{equation}   
 
Here, $\Phi$ is a function of the inclination angle ($i$), the eccentricity ($e$), the longitude of periastron ($\omega$) and the phase of the orbit ($\varphi$), all of which are unknown 
in principle (together with the angular separation):
\begin{multline}
\label{eqn:e2}
\Phi(i,e,\omega,\varphi) = \arrowvert (1-e)(1+\cos{E}) \{ (1-e\cos{E})* \\
[1- \sin^2{(\nu + \omega)} \sin^2 i] \sin{(\nu+\omega)(1+\cos{\nu})\sin{i}} \} ^{-1} \arrowvert
\end{multline}

where the eccentric anomaly E is related to the true anomaly through the relation:
\begin{equation}
\label{eqn:e3}
\tan{\frac{\nu}{2}} = \sqrt{\frac{1+e}{1-e}} \tan{\frac{E}{2}}.
\end{equation}

The orbital phase is linked to the true anomaly $\nu$ and to the eccentric anomaly E through Kepler's equation,
\begin{equation}
\label{eqn:e4}
E-e\sin{E} = 2\pi\varphi
\end{equation}

If the acceleration is measured in m s$^{-1}$ yr$^{-1}$, the angular separation in arcsec, and the distance in pc, the companion mass results in solar units. \\
To apply this method, we computed the slope from the RV data presented in Fig.\ref{fig:plotone}. We excluded from this the CfA DS data, since they have a much 
lower RV precision. For the cases of CD$-$436810 and G135$-$46 we used only data that are on the linear part of the RV curve. For the remaining cases 
(G27-44, G63-5, G237-84, HD16784, HD7424 and HD192718) we used the linear slope of the RV data (summarized for all targets in Table~\ref{table:TOT}), 
which indicates the presence of a distant companion orbiting with a period greatly exceeding the duration of the observations.
We then calculated the distance to the star from the parallax value measured with Hipparcos. Therefore, we have only two quantities, $d(RV)/dt$ and D, 
as observables. The separation, $\rho$, and the orbital elements are unknown. To obtain information on the secondary mass we then followed a numerical Monte Carlo 
approach in which we adopted randomly drawn uniform distributions for the longitude of the periastron (in deg), the phase and a distribution flat in cos $i$ for the inclination angle (in deg). 
We set the orbit to be circular for the case of CD$-$436810, since the solution found from the Keplerian fit is close to circular. 
For the other binaries we set a random eccentricity in the range [0., 0.95]. For the orbital period, we set random, uniformly distributed values up to 100 years, 
while the minimum value was set from the total timespan of the RV measurements. The projected separation was computed using Keplers's third law and using the known Hipparcos 
distance and the randomly set orbital period assuming a total system mass of 1 $M_{\sun}$ (a reasonable approximation).  
From the obtained values of the secondary mass we accepted only values that satisfy the condition $M_{2}$ $<$ $M_{1}$ (these binaries are single-lined spectroscopic and by 
definition the companion star cannot be as massive as the primary, otherwise a secondary spectrum would have been detected in both HARPS and HIRES data). This implies that 
not all values of $\rho$ are accepted. At the end, we can obtain a distribution for the period and the mass of the detected companion (see below).

\subsubsection{An estimate for the upper limit of the minimum mass}
\label{sec:minmass}

The CfA DS data are, of the various RV datasets, the one that has the largest error bars. On the other hand, it spans a period much longer than the one 
obtained with HARPS or HIRES. We can thus use the CfA DS RVs to estimate an upper limits for the minimum mass of the companion, and further constrain the 
range of orbital periods. For this, we fixed K at the value of RMS observed in the RV data (Table \ref{table:RMS}) and used this value to calculate a 
minimum mass (with the Eq.\eqref{eqn:e01}) as a function of orbital period. This upper limit for the minimum mass will be used as an additional constraint 
in Sect. 4 where results for each target are presented.

\subsection{Astrometry}
\label{sec:astrometry}

Whenever possible, we used both the old (\citealt{ESA97}) and new reduction (\citealt{le07}) of Hipparcos catalogue data, 
as well as the proper motion values from Tycho-2 Catalogue (\citealt{hog00}).

\subsubsection{Periodogram}
\label{sec:perio}

We ran a periodogram analysis on the intermediate astrometric data of Hipparcos, following the method used in \citet{soz10}. 
We calculated the chisquare over a large grid of periods, supposing a circular orbit. This was done using both old and new reductions 
of the Hipparcos data, when differences existed. The evaluated fitted model is fully linear in 9 parameters: the five astrometric ones 
and the four Thiele-Innes constants A, B, F and G that represent the orbit of one component around the center of mass \citep{pour00,soz10}. 
We examined the periodograms to find evidence for any short or long -period trends in the data.

\subsubsection{$\Delta\mu$ method}
\label{sec:deltamu}

For every star we also compared the proper motion components, $\mu_{RA}$ and $\mu_{Dec}$, of Hipparcos (\citealt{ESA97, le07}) and Tycho-2 (\citealt{hog00}) and, when possible, 
calculated the difference between them (Table \ref{table:TOT}). This methodology can be used to estimate a value for the mass of the companion, because the Hipparcos catalogue 
includes short-term proper motions only (based on observations collected during a relatively short time), while the Tycho-2 catalogue is based on long-term observations of star 
positions. In binaries of sufficiently long periods, the reflex orbital motion of the primary will be captured in the observed short-term proper motions. The long-term proper motion, 
on the other hand, will be closer to the true center of mass motion of the system. Thus, we define $\Delta\mu$ binaries as stars that have instantaneous (or short-term) 
proper motions significantly different from the quasi-inertial motion of the center of mass (\citet{mak05}). For those stars that have a significant $\Delta\mu$ (in mas yr$^{-1}$) respect more than twice its error.
 
We used the following expression, described in \citet{mak05}, to estimate the mass of the companion $M_{2}$:
\begin{equation}
\label{eqn:e5}
\Delta\mu \leq \frac{2\pi \Pi R_{0} M_{2}}{M_{tot}^{2/3} P^{1/3}},
\end{equation}

where $\pi$ is the parallax in mas, $P$ the orbital period in years, $M_{2}$ is the secondary mass, and $M_{tot}$ is the total mass of the system 
(both in $M_{\sun}$). The $R_{0}$ parameter is an orbital phase factor that we considered constant and equal to 1, under the assumption of circular orbits. 
We calculated $\Delta\mu$ for a range of combination of periods (in days) and masses of the companion and used Eq.\eqref{eqn:e5} to valuate the possible 
solutions for the system (accepting only values for the mass of the companion that satisfy the condition $M_{2}$ $<$ $M_{1}$). \\ 

\subsubsection{$\dot{\mu}$ method}
\label{sec:accel}

Owing to their orbital motion around a companion star, some stars cannot be accurately described in the Hipparcos astrometry 
by a five-parameter model that includes two position components, two components of proper motion, and parallax as unknowns. 
Therefore, a more complex model of seven free parameters (including the acceleration of the proper motion components -- $\dot{\mu}$) may be necessary. 
About 2.2 $\%$ of stars in the $Hipparcos$ catalogue require such a special treatment (\citet{per97}).  \\
\citet{mak05} collected all currently known $\dot{\mu}$ binaries in the Hipparcos catalogue. The following relation was then used 
to estimate the expected $\dot{\mu}$ (mas yr$^{-2}$) as a function of a range of orbital periods (days) and masses for the companion ($M_{\sun}$: 
 
\begin{equation}
\label{eqn:e6}
\dot{\mu} \leq \frac{(2\pi R_{1})^2 \Pi M_{2}}{M_{tot}^{2/3} P^{4/3}}
\end{equation} 

Accepted values were those satisfying the condition $M_{2}$ $<$ $M_{1}$) that provide a value higher or equal to the observed $\dot{\mu}$.
Under the assumption of circular orbits we kept $R_{1}$, which is the orbital phase factor, constant and equal to 1. \\
Unfortunately, in the Hipparcos data an acceleration solution is available only for one of our targets (CD$-$436810). This is due to a combination of long 
orbital periods and large distances for the binary sample investigated here.

\section{Results}
\label{sec:results}

\subsection{Case by case analysis}
\label{sec:analysis}
 
In this section we present the results of our case by case analysis, using the methodology steps presented above.

\subsubsection{HD16784}
\label{sec:HD16784}

\begin{figure*}[p]
  
  \begin{tabular}{cc c}
    \includegraphics[width=8cm]{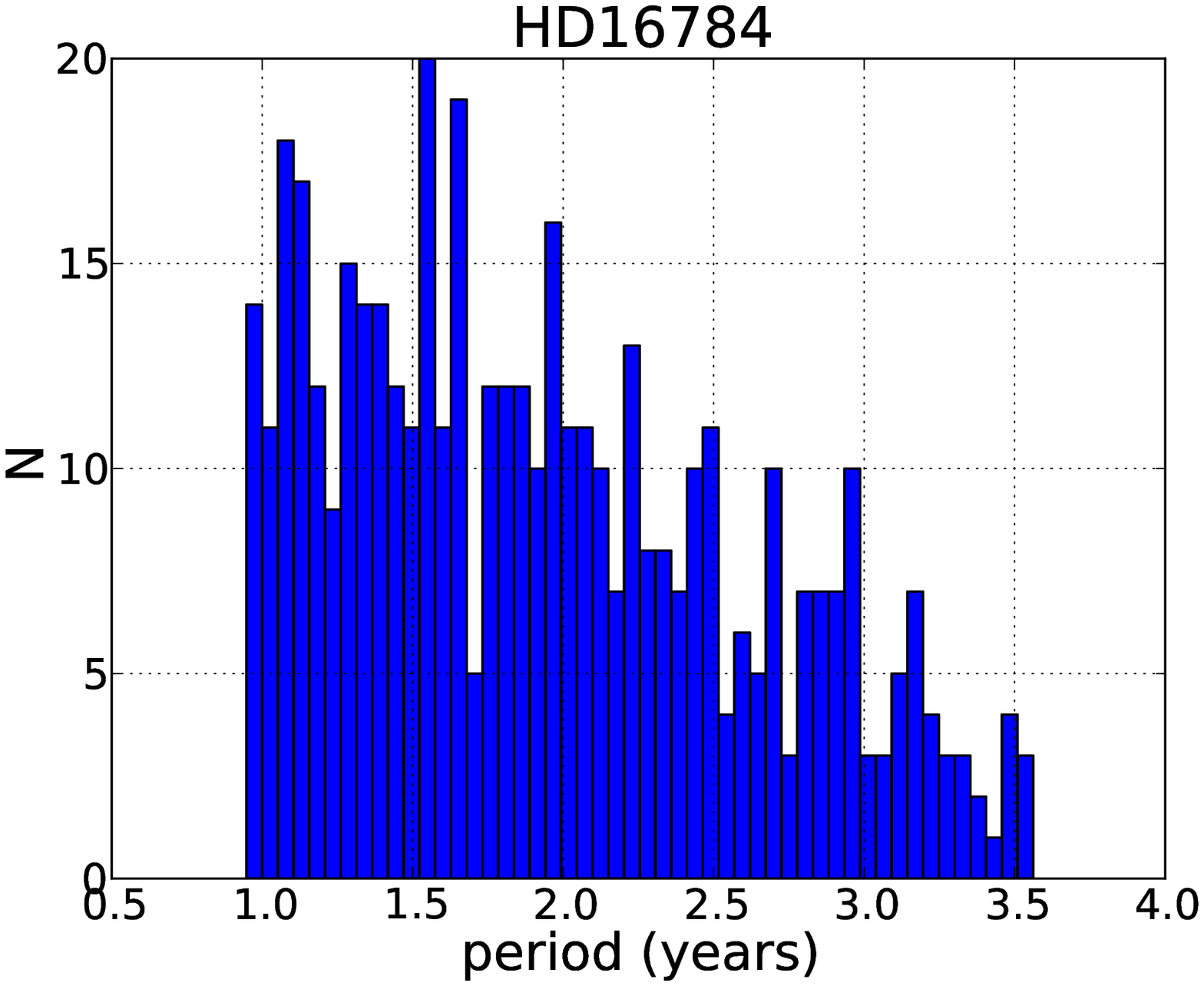}&

    \includegraphics[width=8cm]{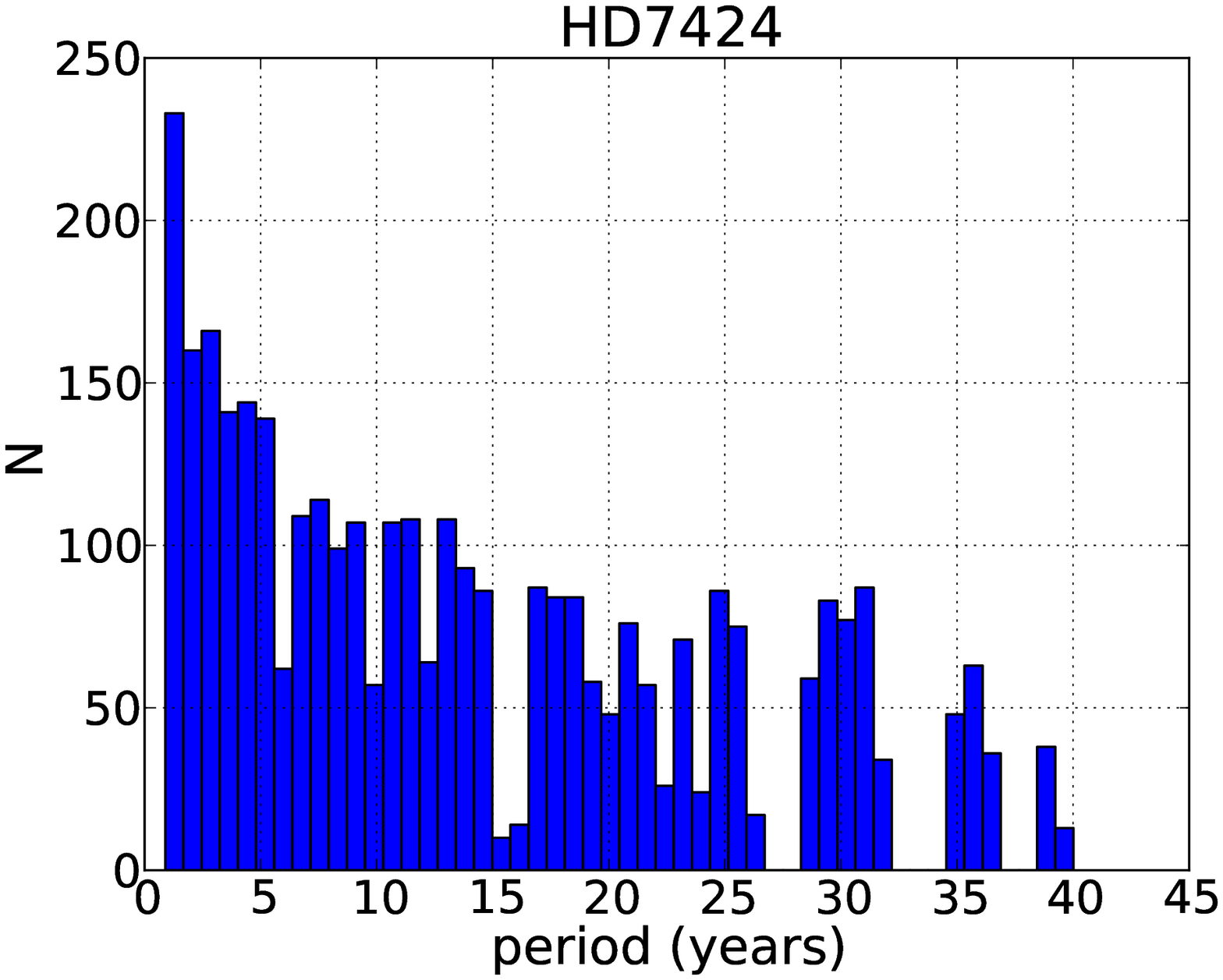}\\

    \includegraphics[width=8cm]{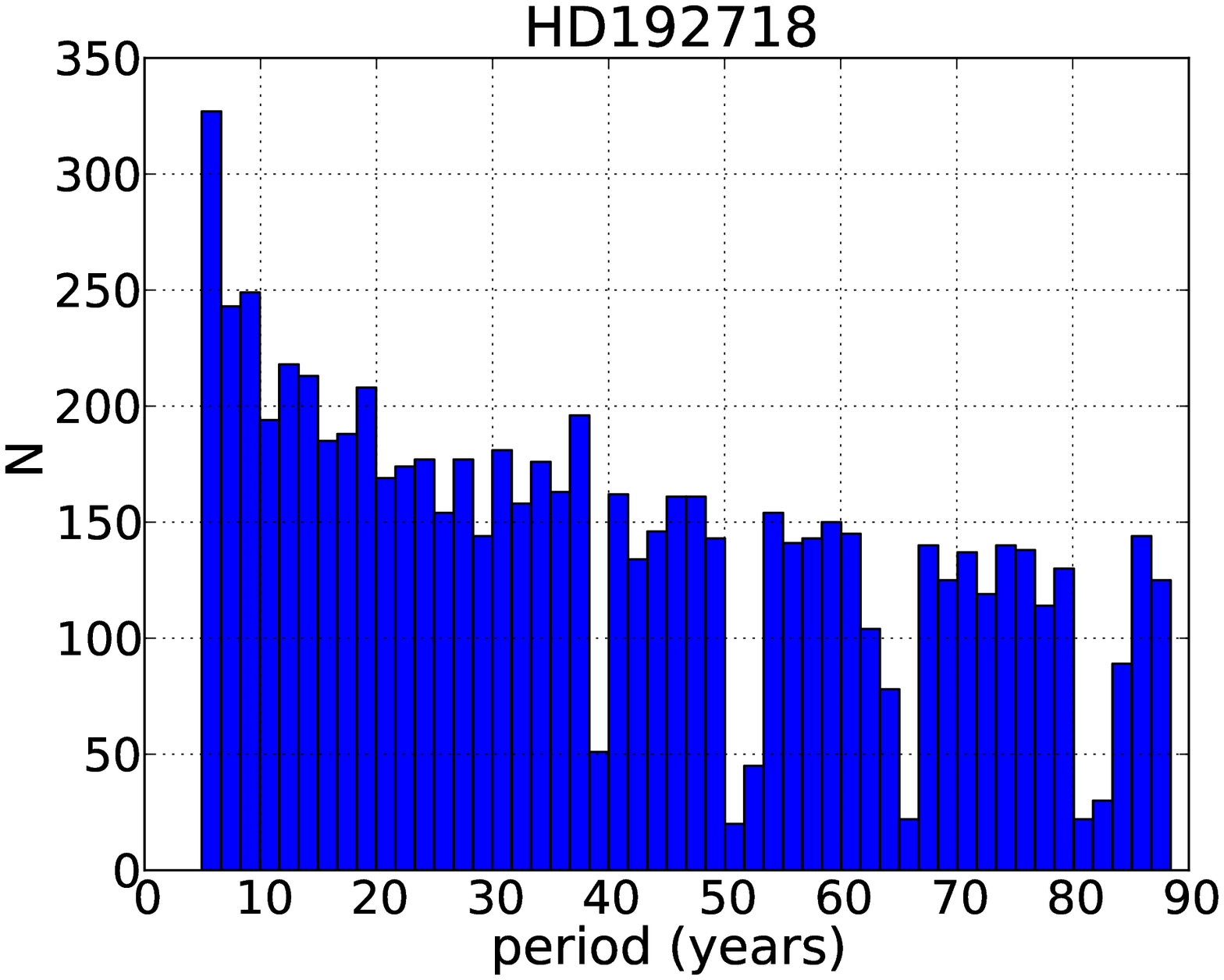}
         
   \end{tabular}
   \caption{Limits on the orbital period to the companions to HD16784 (\emph{upper left}),HD7424 (\emph{upper right}) and HD192718 (\emph{bottom}).}\label{fig:plottwo}
   \end{figure*}

For this case, we used only the method of \citet{tor99}, using the slope (10525.488 $\pm$ 6169.67 m s$^{-1}$ yr$^{-1}$) 
of the RV measurements (Nmes = 3). As seen in Fig. \ref{fig:plotone}, the limited number of measurements and the actual non-linearity of the RV trend translate in a not well-constrained RV slope.   
The timespan is $<1$ years (the data cover 272 days). From the analysis, for the mass the histogram does not show a very definite 
result but the mass of the companion seems to be $>$ 0.13 $M_{\sun}$. The orbital period results between 1 year and below 3 years and half with 
1 $\sigma$ (68 $\%$) at 2.2 years and 2 $\sigma$ (95 $\%$) at 3.2 years (Fig. \ref{fig:plottwo}).
In this case, we did not use the $\Delta\mu$ method since no significant $\Delta\mu$ is observed (see Table \ref{table:TOT}). Moreover, we did not find any signal doing the periodogram. 
No $\dot{\mu}$ is available for this star.

\subsubsection{G27-44}
\label{sec:G27-44}

\begin{figure*}[p]
  
  \begin{tabular}{cc c}
    \includegraphics[width=8cm]{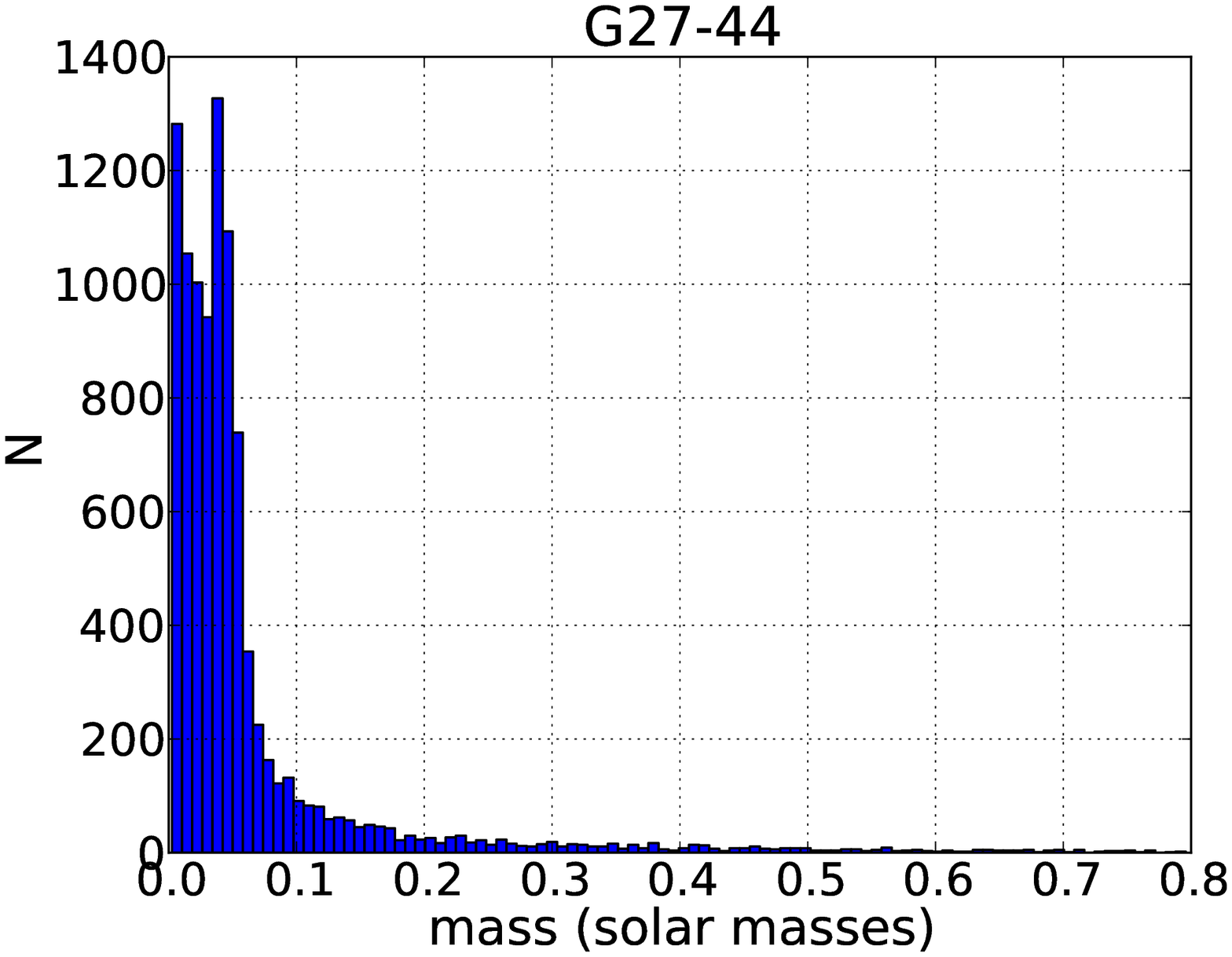}&

    \includegraphics[width=8cm]{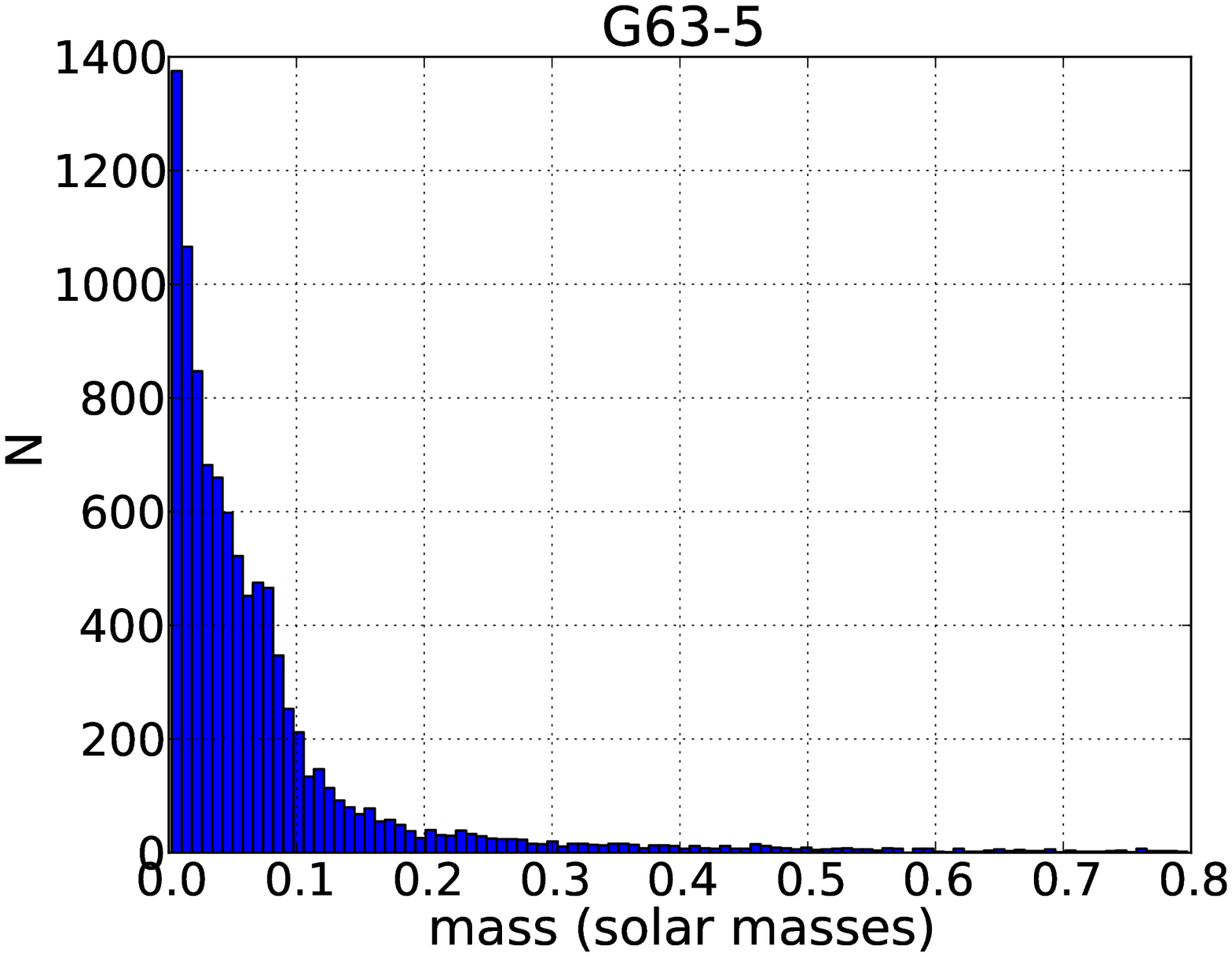}\\

    \includegraphics[width=8cm]{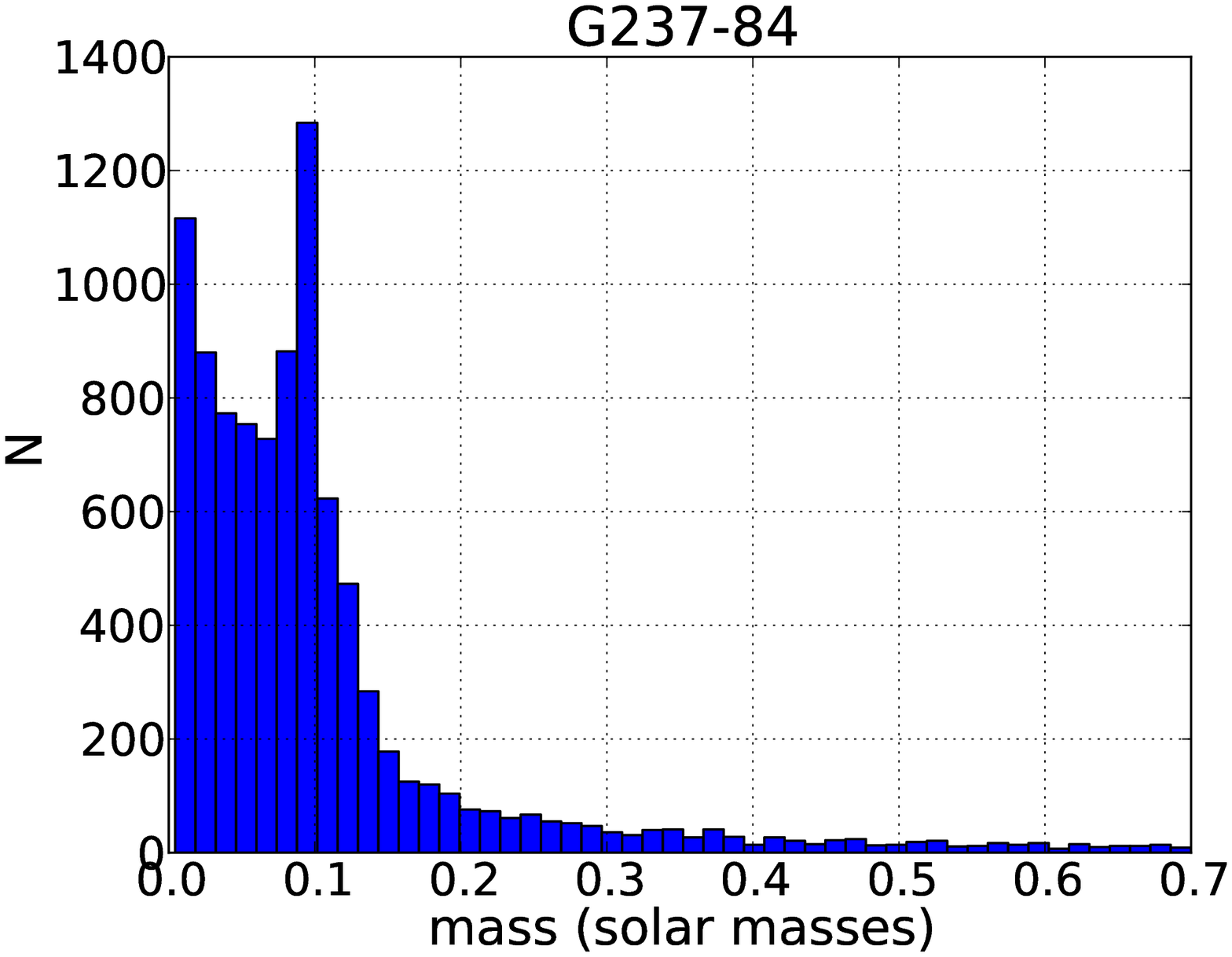}
       
   \end{tabular}
   \caption{Limits on the mass to the companions to G27$-$44 (\emph{upper left}), G63$-$5 (\emph{upper right}) and G237$-$84 (\emph{bottom})}\label{fig:plotthree}
   \end{figure*}

The radial velocities (Nmes=23) of this stars suggest that the period is longer than $\sim$ 8 years because the data has a total timespan of 3200 days, 
but still the radial velocity curve is not complete. The RV data for this system shows a clear deviation from a straight line model 
(as can be seen from a visual inspection of Fig.\ref{fig:plotone}). 
Using the method of \citet{tor99}, we analyzed the slope ($-$11.20 $\pm$ 0.78 m s$^{-1}$ yr$^{-1}$) of the RV data to estimate the mass of the companion and its period. 
Although there is a significant acceleration, due to the small number of measurements we decided to use all the data in this analysis. 
The analysis did not provide any clear conclusion regarding the orbital period. The solutions found suggest that, with a 1-$\sigma$ confidence, 
it should be shorter than 61 years. Note, however, that as mentioned above, we limited our analysis to periods up to 100 years. On the other hand, 
we find a very interesting result concerning the mass of the companion. From the results shown in Fig.\ref{fig:plotthree} we find that at a 1-$\sigma$ confidence level 
the mass of the companion is $<$ 0.05 $M_{\sun}$ ($M_{\sun}$$<$0.25 at  2-$\sigma$) and the lowest allowed value of the secondary mass from the histogram is 0.002 $M_{\sun}$. 
From the analysis of the CfA DS data (RMS $=$ 0.39 km s$^{-1}$) we also conclude that the upper limit of the $M\sin{i}$ is $~$0.03 $M_{\sun}$ for a period of $\sim$16 years (the time span of the CfA DS data). 
In this case, we did not use the $\Delta\mu$ method since no significant $\Delta\mu$ is observed (see Table \ref{table:TOT}). 
Moreover, no trend in the Hipparcos data, doing the periodogram, appears. No $\dot{\mu}$ is available for this star.

\subsubsection{G63-5}
\label{sec:G63-5}

The 16 measurements for this star were taken during a time span of $\sim$7 years (2700 days), presenting only a straight linear trend. 
Analyzing this slope (12.37 $\pm$ 1.07 $m s^{-1} yr^{-1}$) we note that also in this case we can find, at 68 $\%$ and 95 $\%$ of confidence levels, a mass for the companion $<$ 0.07 $M_{\sun}$ and 0.28 $M_{\sun}$, respectively (Fig. \ref{fig:plotthree}) and with a minimum mass of 0.002 $M_{\sun}$. 
For the period, the histogram does not show a very definite result. 
Calculating the maximum value for the minimum mass (RMS $=$ 0.36 km s$^{-1}$), we obtain an upper limit of 0.025 $M_{\sun}$ for a period of $\sim$8\,years (the span of the TRES RV data). 
Again for this case, we did not use the $\Delta\mu$ method since no significant $\Delta\mu$ is observed (see Table \ref{table:TOT}). 
Moreover, no trend in the Hipparcos data, doing the periodogram, appears.
No $\dot{\mu}$ is also available for this star.

\subsubsection{G237-84}
\label{sec:G237-84}

For this star we have a significant value for $\Delta\mu$ in addition to the slope (23.58 $\pm$ 1.72 $m s^{-1} yr^{-1}$) of the RV data (Nmes=38).
With the slope information we find at 68 $\%$ the mass $<$ 0.10 $M_{\sun}$ and at 95 $\%$ $<$ 0.35 $M_{\sun}$ (Fig. \ref{fig:plotthree}) and the minimum mass of the companion found in the 
histogram is 0.004 $M_{\sun}$. The period is found to be $>$ 6 years (2300 days). From the period's histogram we cannot take any firm conclusions. 
From the study of the maximum minimum mass (RMS $=$ 0.73 km s$^{-1}$) we derive a value $<$ 0.07 $M_{\sun}$.
For this star the $\Delta\mu$ calculated in the literature has a significant value, but did not allow us to add some further constraints for the mass of the companion 
(depending on its orbital period). The value of $\Delta\mu$ used in this case (Table \ref{table:TOT}) corresponds to the right-ascention component of the proper motion (the larger one). 
No $\dot{\mu}$ is available and no trend in the Hipparcos data, doing the periodogram, appears for this star.

\subsubsection{HD7424}
\label{sec:HD7424}

For this star we could use both the $\Delta\mu$ and the slope (-417.02 $\pm$ 4.66 $m s^{-1} yr^{-1}$ in 500 days) methods. 
We didn't find very strong constraints for the mass from the $\Delta\mu$ analysis. 
From the slope analysis we find interesting limits: the mass results to be $>$ 0.008 $M_{\sun}$ and at 1 $\sigma$ $<$ 0.43 $M_{\sun}$.
For the period we find strong constraints when compared with the cases discussed in the previous sections. A value between 1 and 40 years, 
with a maximum of 19-years at a 68 $\%$ (1-$\sigma$) confidence level (Fig. \ref{fig:plottwo}). No allowed values above 40 years exist due to the companion 
upper mass constraint used in our simulations.
The analysis of the dispersion of the CfA DS data (RMS $=$ 0.51 km s$^{-1}$) suggest that the maximum value for the minimum mass of the companion is 0.07 $M_{\sun}$.
No trend in the Hipparcos data, doing the periodogram, appears in this case.
Moreover, no $\dot{\mu}$ is available for this star.

\subsubsection{HD192718}
\label{sec:HD192718}

For this case we used only the method of \citet{tor99} having the slope (-144.65 $\pm$ 1.40 m s$^{-1}$ yr$^{-1}$) of the RV measurements (Nmes=20). 
The period is longer than 5 years (the data cover 2200 days) and from the analysis we can note that the mass seems to be $>$ 0.02 $M_{\sun}$, 
while at 1 and 2 $\sigma$ confidence levels it is $<$ 0.55 and 0.75 $M_{\sun}$, respectively. 
The obtained period is $<$ 53 years at 1 $\sigma$ and $<$ 83 years at 2 $\sigma$ (Fig. \ref{fig:plottwo}). 
From the analysis of the CfA DS data (RMS $=$ 0.92 km s$^{-1}$) we also conclude that the upper limit for the minimum mass of the companion is 0.08 $M_{\sun}$. 
No $\dot{\mu}$ data of is available for this star, the $\Delta\mu$ does not show any significant value and no trend in the Hipparcos data, doing the periodogram, appears.  
Taking into account the TRES data we observe that the orbital period is expected to be longer than $\sim$ 13.5 years (5000 days) and results from a Keplerian fit are compatible with the analysis, 
with an orbital period around 46.5 years ($\sim$ 17000 days).

\subsubsection{G135-46}
\label{sec:G135-46}

\begin{figure*}[p]
   \centering

  \begin{tabular}{cc cc}
    \includegraphics[width=8cm]{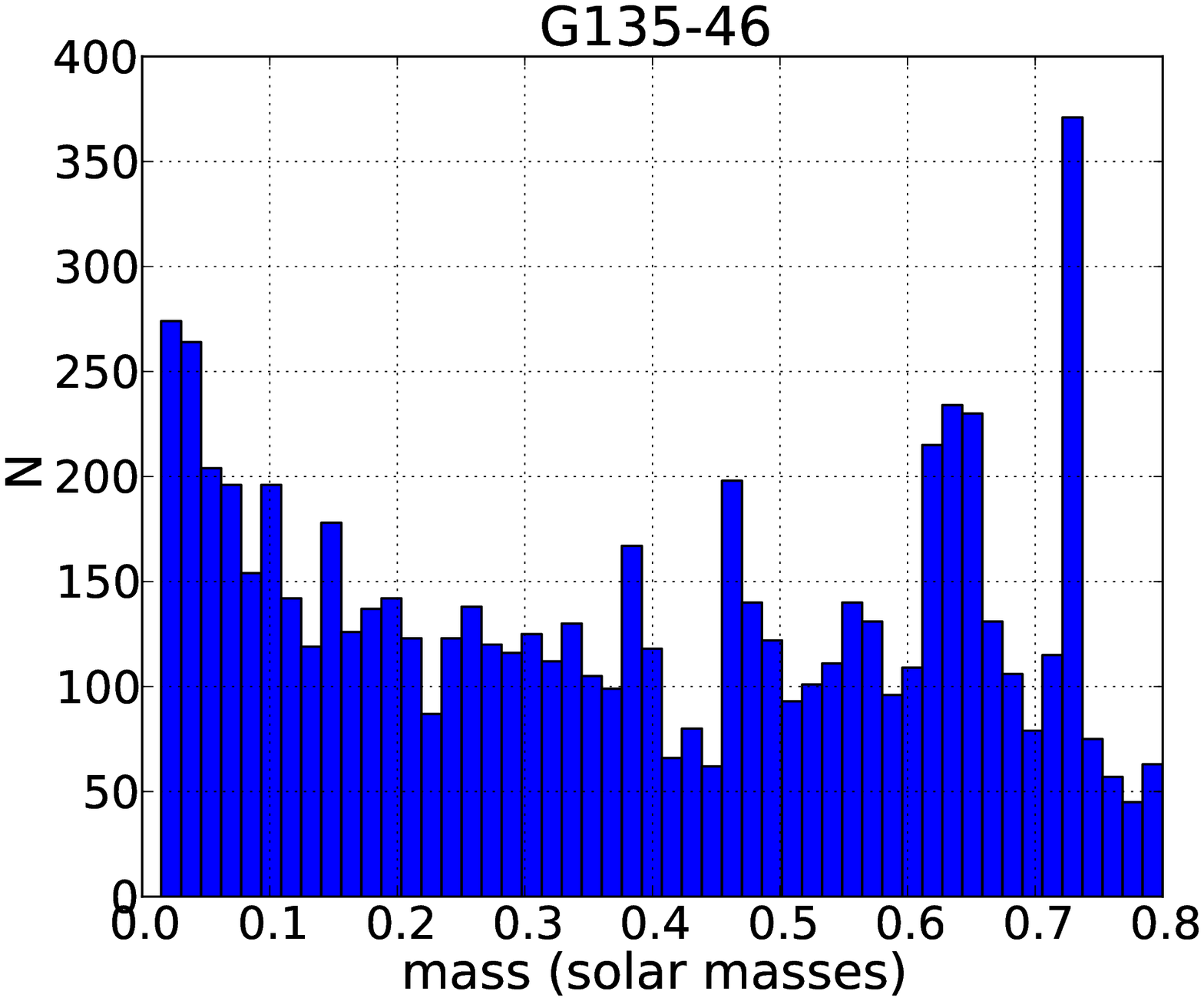}&

    \includegraphics[width=8cm]{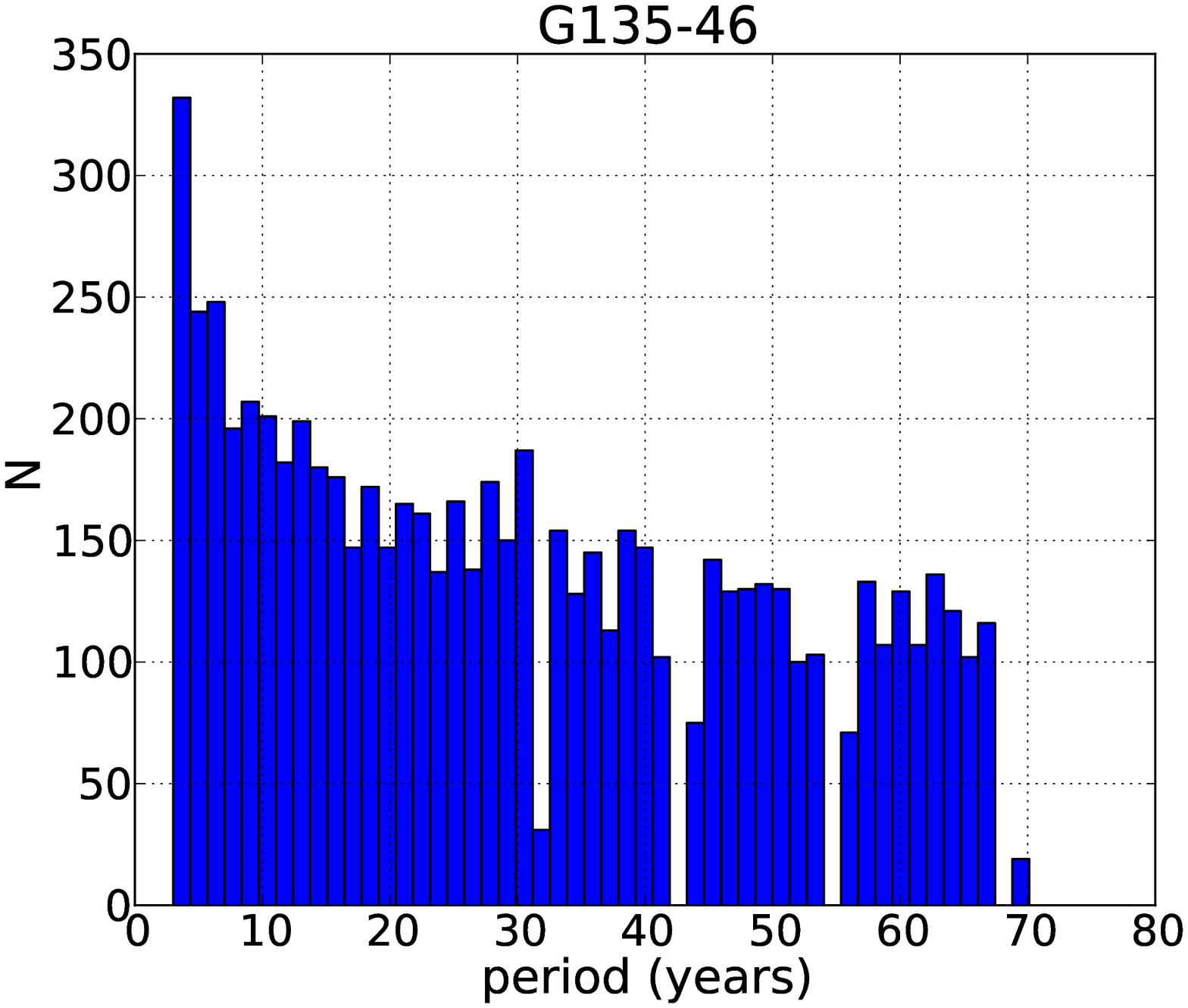}\\

    \includegraphics[width=8cm]{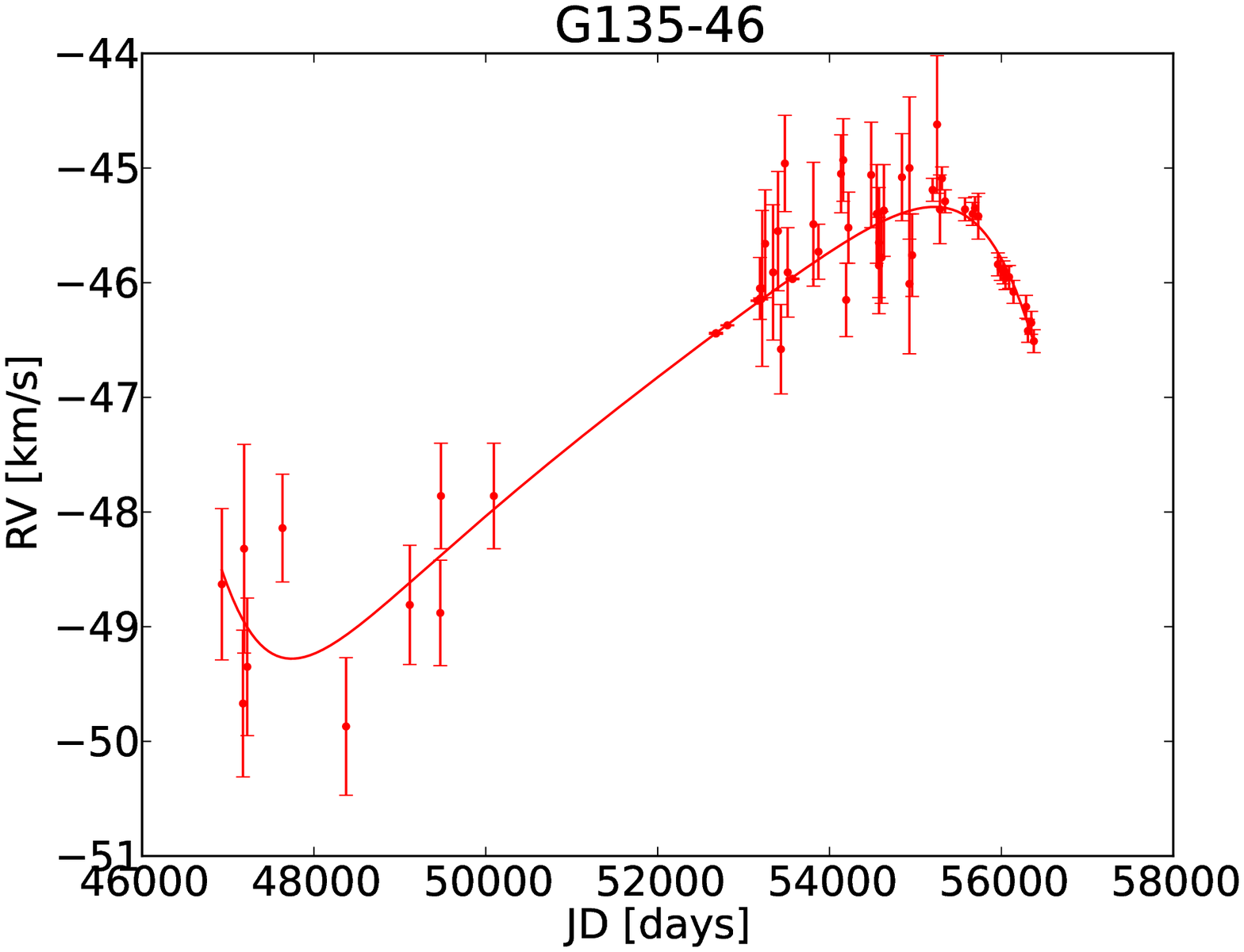}&

    \includegraphics[width=8cm]{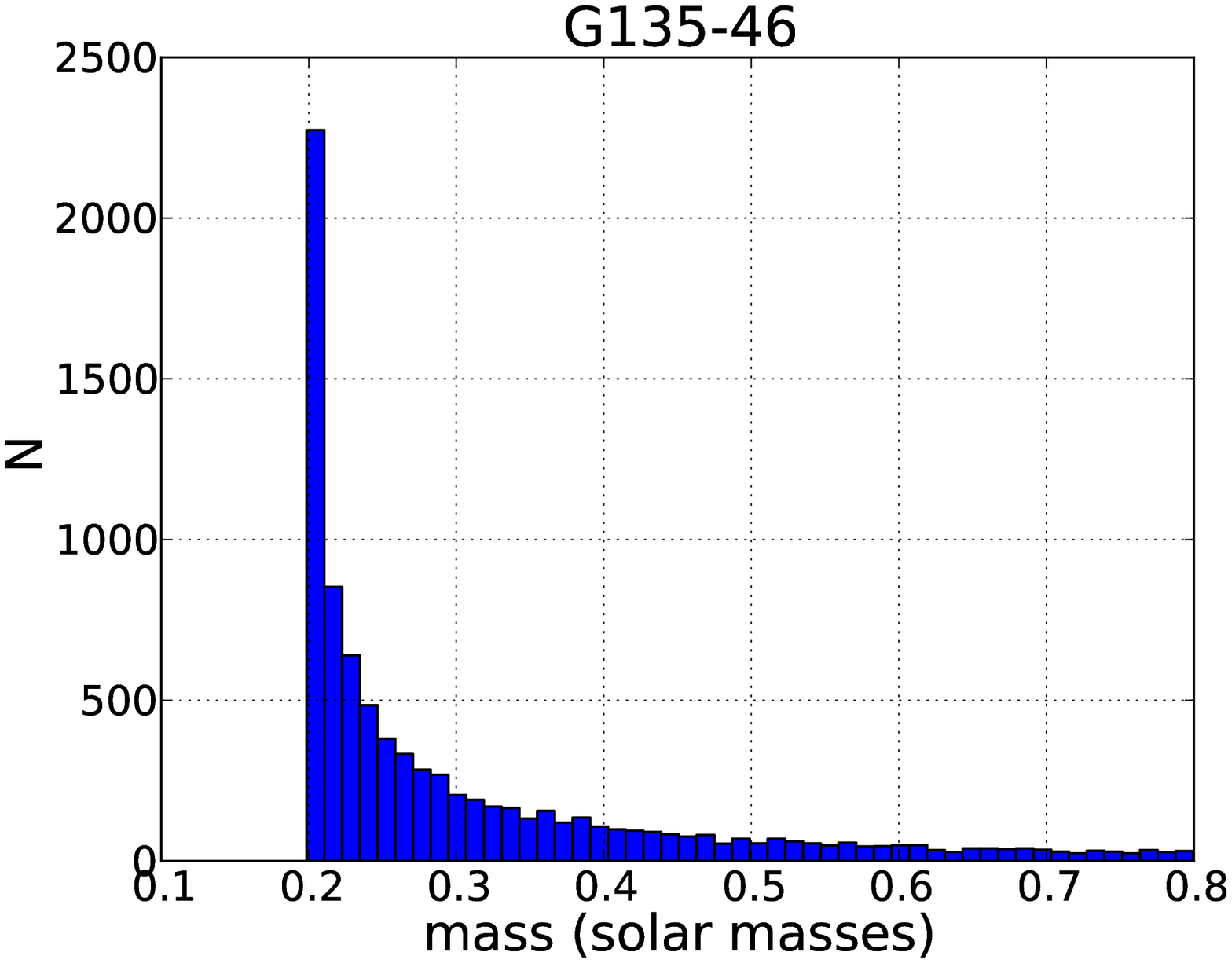}\\
         
   \end{tabular}
   \caption{\emph{Upper two panels}: Limits on the mass and the orbital period of the companion to G135$-$46. 
   \emph{Bottom left}: the RV data with overplotted the best-fit Keplerian orbit. The last plot (\emph{right panel}) 
   is the histogram of the mass fixing the period at the orbital fit value.}\label{fig:plotfour}
   \end{figure*}

Thanks to the CfA DS and TRES data for this binary we could almost complete one cycle in the RV curve. Using only the slope of the Keck data 
(Nmes=9 in 900 days) and the Torres method we could not find any strong constraints for the mass of the companion and for the orbital period of the system 
In fact, from the histogram in Fig. \ref{fig:plotfour} (\emph{left panels}), we can only infer that the mass of the companion is $<$ 0.54 $M_{\sun}$ at 68 $\%$ 
(Fig \ref{fig:plotfour}) and $<$ 0.73 at 95 $\%$. The orbital period's histogram shows, on the other hand, that the period is shorter than 70 years, 
while at 1 and at 2 $\sigma$ levels its value is 39 years and 63 years, respectively (Fig \ref{fig:plotfour}). 
The new data we could however fit the RV curve and derive the orbital properties of the binary system and the minimum mass of the companion. 
The results are shown in Table \ref{table:4}. The minimum mass results to be 0.2 $M_{\sun}$. 
With this value, we can estimate again the mass of the companion using the Torres method but imposing the period of 27 years resulting from the 
Keplerian fit to the Doppler measurements and the actual mass results to be $>$ 0.2 $M_{\sun}$. 
With this extra constraint, we derive at 1 and 2 $\sigma$ limits a mass of $<$0.30 $M_{\sun}$ and $<$0.63 $M_{\sun}$, respectively.
We did not use the $\Delta\mu$ method since no significant value is observed (see Table \ref{table:TOT}).
Moreover, no trend in the Hipparcos data, doing the periodogram, appears. 
No $\dot{\mu}$ is also available for this star.

\subsubsection{CD-436810}
\label{sec:CD-436810}

\begin{figure*}[p]
   \centering

  \begin{tabular}{cc cc}
    \includegraphics[width=8cm]{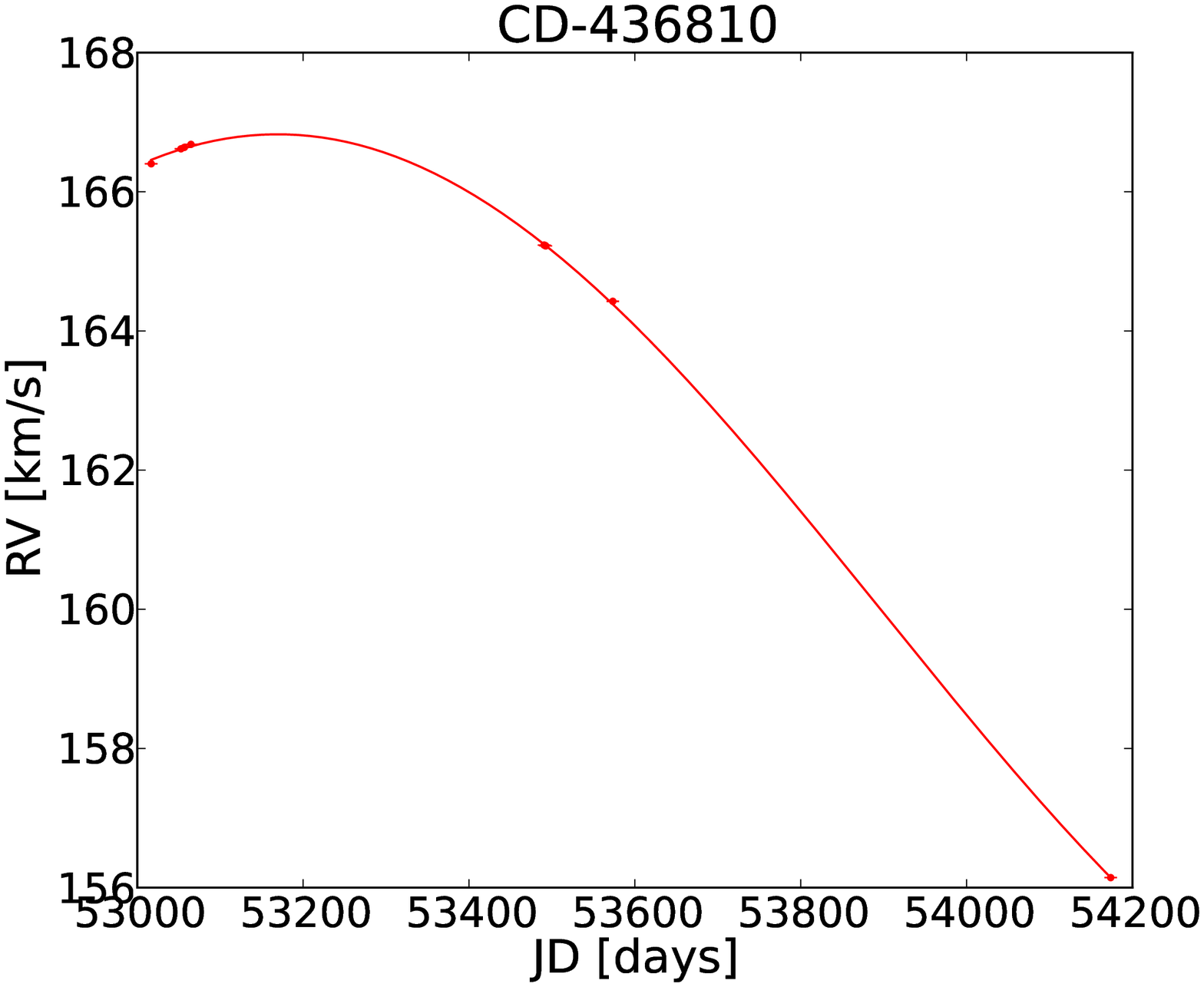}&

    \includegraphics[width=8cm]{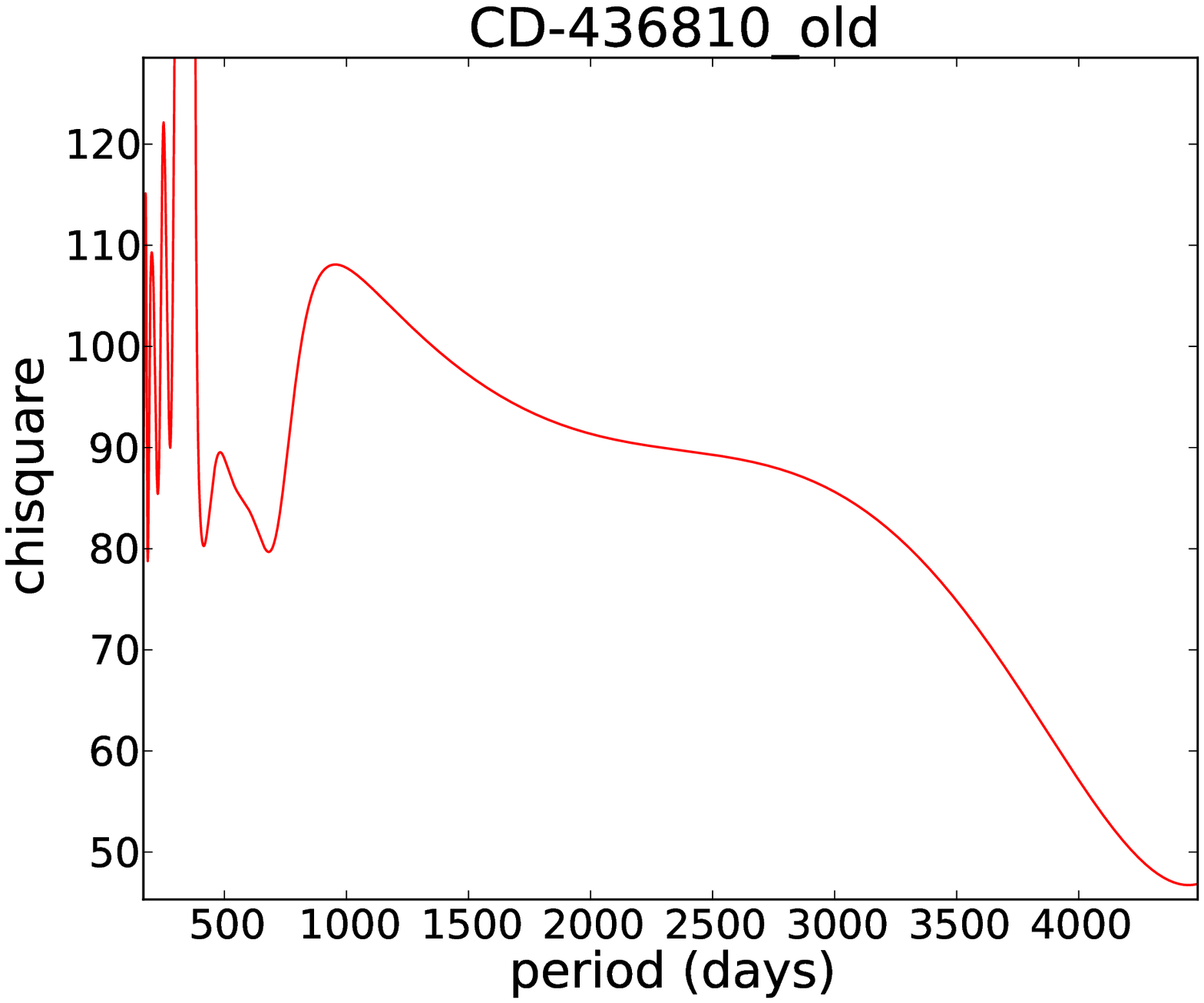}\\

    \includegraphics[width=8cm]{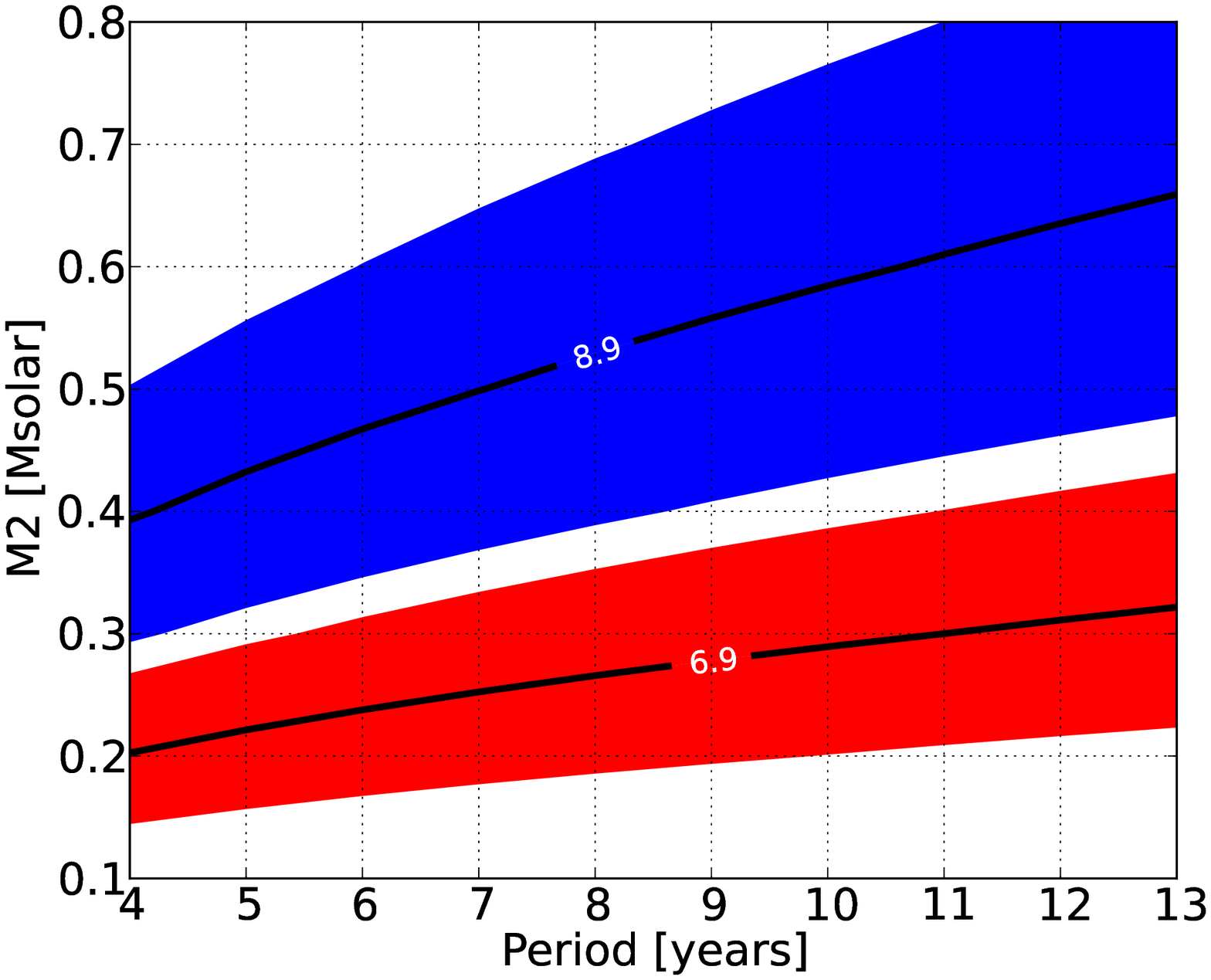}&

    \includegraphics[width=8cm]{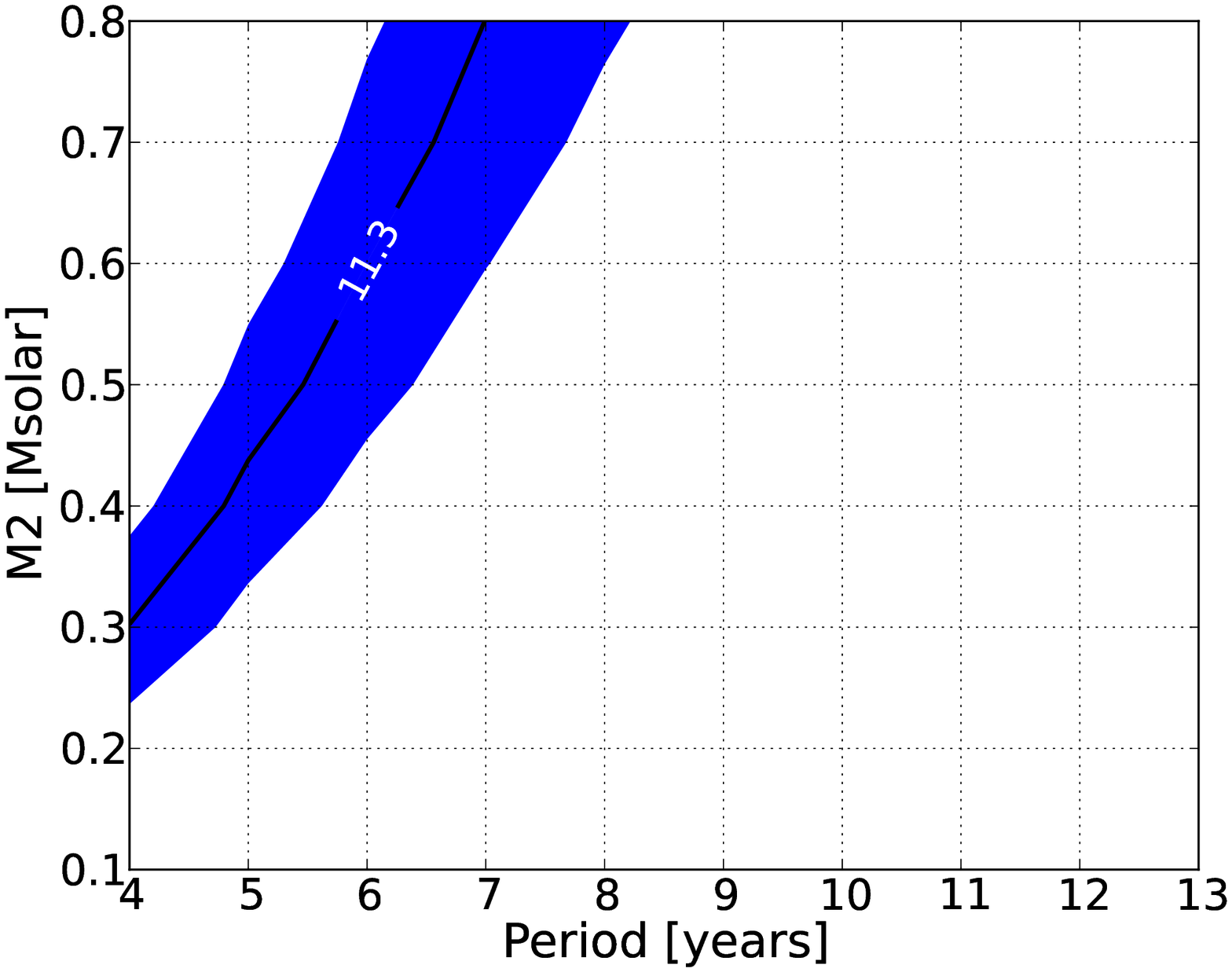}\\
         
   \end{tabular}
   \caption{\emph{Upper left:} The Keplerian orbital fit to the CD-436810 RV data.
\emph{Upper right and bottom:} The periodogram analysis of the Hipparcos old reduction data and the study of $\Delta\mu$ and $\dot{\mu}$ for the case of CD$-$436810. 
   The results for the old Hipparcos reduction are in blue (with 1 $\sigma$ of confidence level), while in red are reported the results for the new reduction of Hipparcos data. 
   The black lines represent the measured values of $\Delta\mu$ and $\dot{\mu}$.}\label{fig:plotfive}
   \end{figure*}

\begin{figure*}[p]
  \centering

  \begin{tabular}{cc cc}

    \includegraphics[width=8cm]{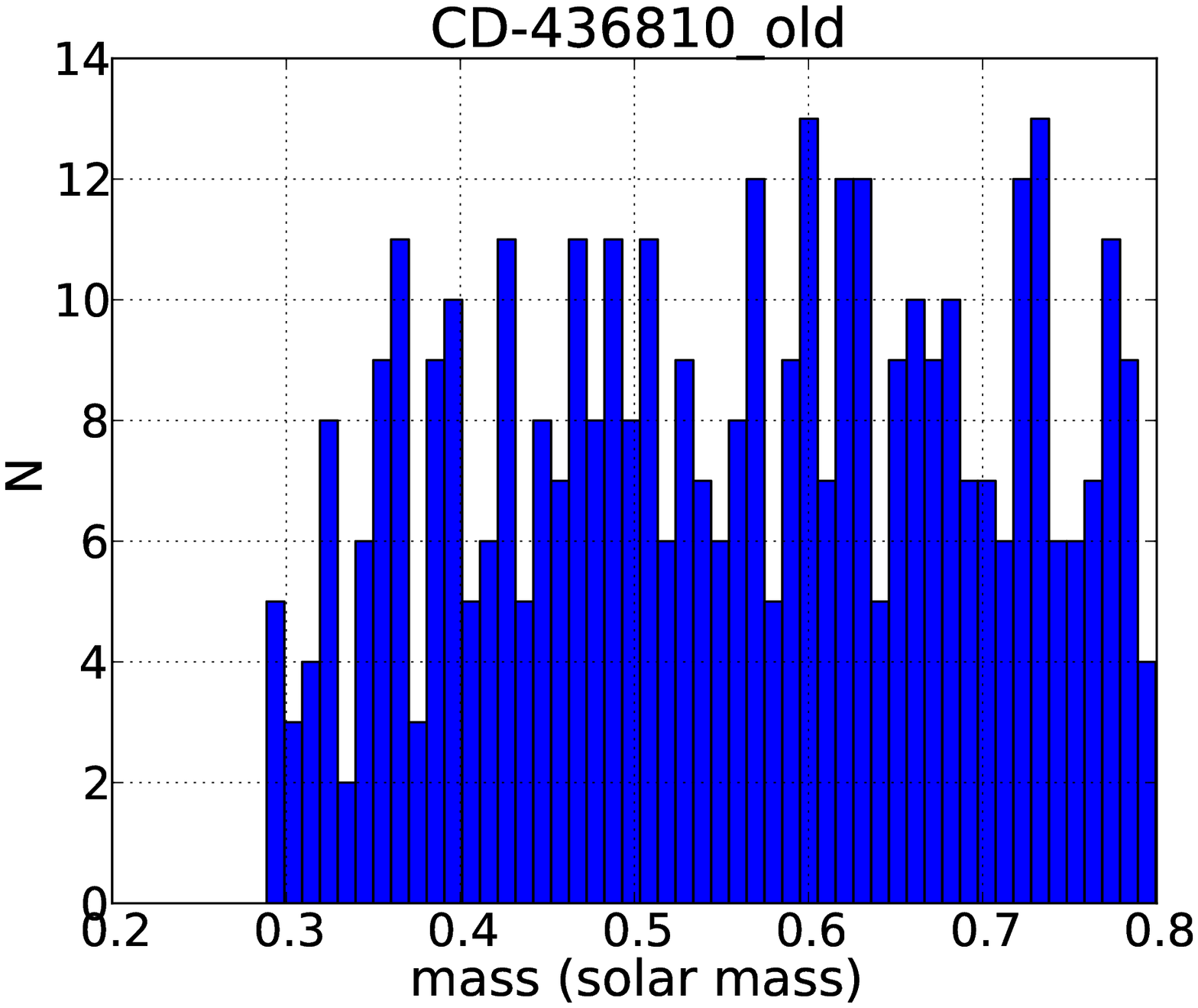}&

    \includegraphics[width=8cm]{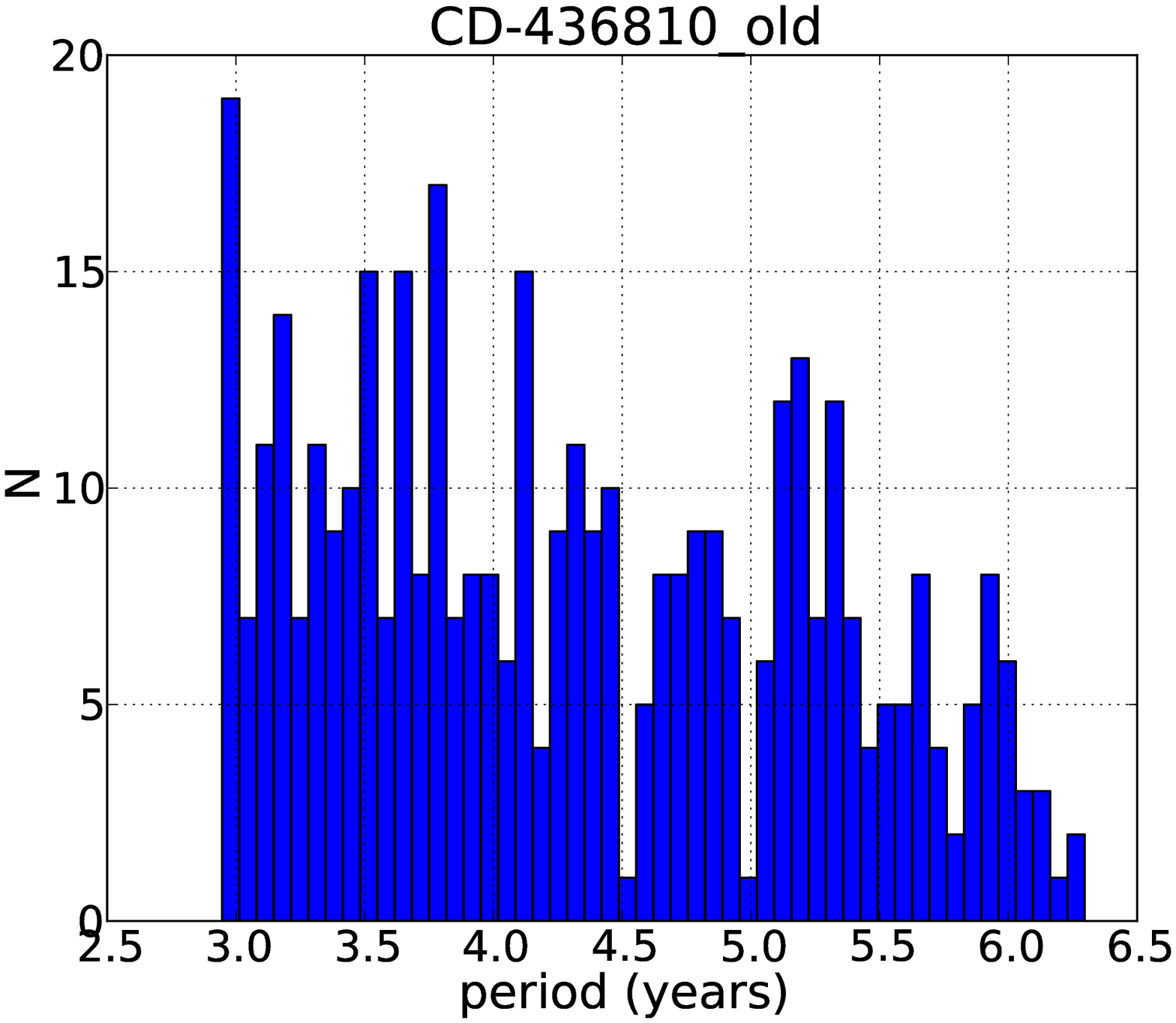}\\

    \includegraphics[width=8cm]{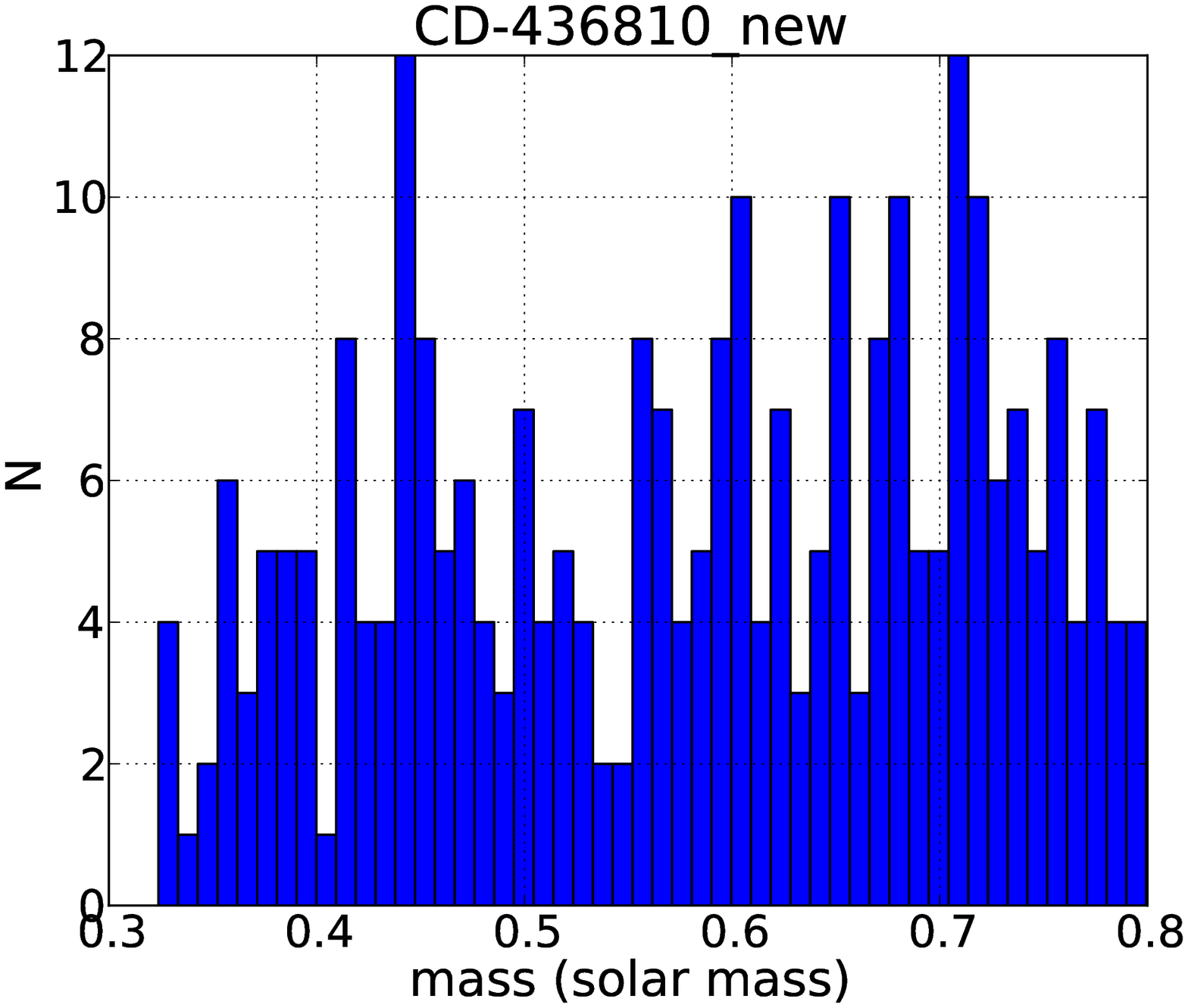}&

    \includegraphics[width=8cm]{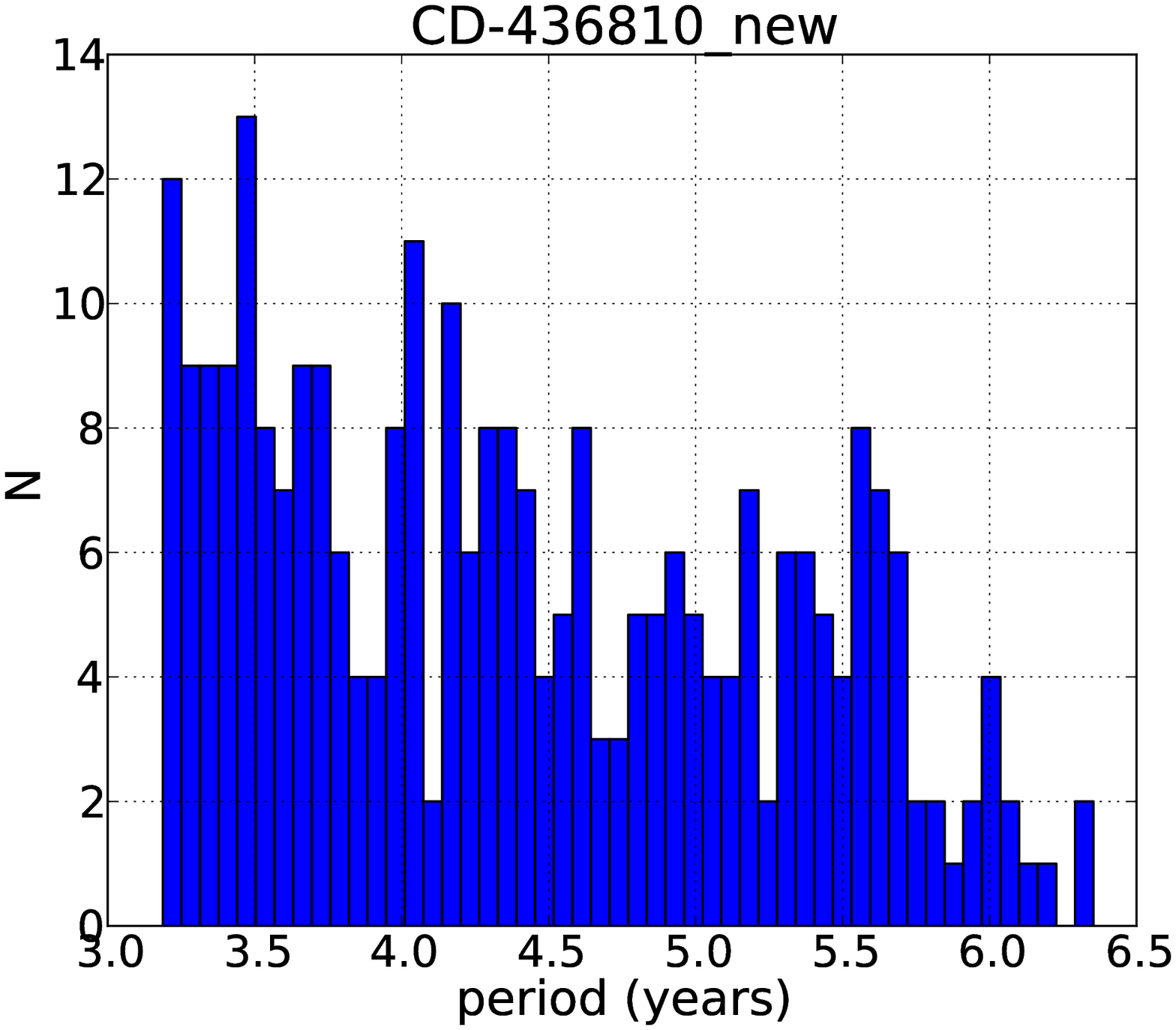}\\

  \end{tabular}
  
  \caption{\emph{Upper panels}: Limits on the mass and the orbital period of the companion to CD-436810 using the old Hipparcos reduction. 
  Limits on the mass and the orbital period of the companion to CD-436810 based on the new Hipparcos reduction are shown in bottom two panels.}\label{fig:plotsix}
\end{figure*}

Also for this star it was possible to put more stringent limits on the companion characteristics. 
For this binary we had RV data (even if not complete, Nmes=9 with a timespan of 3 years and half), significant values of $\Delta\mu$ and $\dot{\mu}$. 
This star is also interesting because it has two contrasting results between the old and the new reduction of the Hipparcos Intermediate Astrometric Data (IAD), respectively. 
The solution found with 7 (old reduction) and 5 (new reduction) parameters provides different parallax values: 6.83 mas for the old and 9.27 mas for the new reduction. 
A periodogram analysis, following the method described in \citet{soz10}, on the Hipparcos data highlighted the presence of a significant long-period trend (Fig.~\ref{fig:plotfive}). We found, based on the F$-$test, that the addition of four parameters to the model of the Hipparcos IAD improved significantly the fit: P(F) = 0.0005. 
This star is the only one for which a periodogram analysis provided corroborating evidence in the Hipparcos data of the presence of a wide-separation companion. 
The parameters of a keplerian fit to the data are shown in the Table~\ref{table:4}. The obtained orbital period is $\sim$ 1702 days ($\sim$ 4.6 years), 
while the derived minimum mass for the companion is $\sim$ 0.28 $M_{\sun}$. However, given that the full orbit is not covered, the solution found may not be optimal and, therefore, it is used as one consistency checks in support of the attempt at constraining the mass and orbital period of the companion to this particular star. 
We thus decided to add more constraints using the other methods mentioned previously in this paper, and making use of the available astrometric data.
For the old Hipparcos reduction, the observed $\Delta\mu$ shows that the mass of the companion is above 0.3 $M_{\sun}$. 
With the value of $\dot{\mu}$ we also find that the orbital period is between 4 - 8 years ($\sim$ 1500 - 3000 days) (Fig. \ref{fig:plotfive}), 
since higher values would imply companion masses above 0.8 M$_{\sun}$. \\
With the new Hipparcos reduction we found a possible lower mass. In Fig. \ref{fig:plotfive} it is possible to observe the comparison between the two reduction for the $\Delta\mu$ and note the difference in the solution.\\
Calculating the mass of the companion and the period with the method of \citet{tor99} we find also different results depending on whether 
we use the old or new reduction parallax results. For the old reduction, the period results between 2.9 years and below 6.3 years with the the 1 $\sigma$ threshold at 4.8 years (Fig. \ref{fig:plotsix}). For this case, the derived secondary mass is $>$ 0.29 $M_{\sun}$ and $<$ 0.80 $M_{\sun}$ with the 68 $\%$ confidence limit being $<$ 0.63 $M_{\sun}$ (Fig. \ref{fig:plotsix}). Using the parallax value from the new reduction, the possible values for the period slightly increase to between 3.2 years and 6.3 years with $P<$ 4.8 years at 1 $\sigma$ confidence level (Fig. \ref{fig:plotsix}). 
The companion mass also changes to values between 0.32 $M_{\sun}$ and 0.80 $M_{\sun}$, with $M_{2}$ $<$ 0.67 $M_{\sun}$ at 68 $\%$ confidence (Fig. \ref{fig:plotsix}). 
Table \ref{table:CD} shows the summary of the results for this binary system.

\section{Discussion and conclusions}
\label{sec:discus}

In this paper we have presented a multi-technique analysis of 8 metal-poor SB1 binary systems with no previously determined orbital solutions and companion mass estimates. 
We found a range of solutions for the secondary masses and for the orbital period of the systems. 
For the examined systems we can conclude that for 3 of them the most likely values for companion mass are below 0.2 solar masses. Two of these systems (G27-44 and G63-5) 
show a possible mass (at a 1 $\sigma$ of confidence level) of  $\approx0.07$ solar masses, putting them as candidate brown dwarf companions. 
For the other 5 binaries we have 4 systems (HD192718, G135-46, HD16784 and CD-436810) where it is more likely that the mass of the companion is higher and in particular 
(for the last) above 0.28 $M_{\sun}$. These stars are thus likely orbited by M dwarf companions.\\ 
A particular consideration is worth for the case of CD-436810, for 
which we find a very interesting result using the two different reductions of the Hipparcos data. 
It seems that the results found using the information in the slope of the RV data are more compatible with the $\Delta\mu$ and the presence of the acceleration solution ($\dot{\mu}$) 
using the old Hipparcos reduction.
The resuls of $\dot{\mu}$ agree with the other limits derived using the other methods. The new reduction find a parallax 25$\%$ higher. Therefore, the perturbation due to the companion should be more evident. Rather than a case of spurious acceleration (e.g., \citealt{tok12, tok13}), this may point to a problem with the new Hipparcos data reduction for this specific star.\\
This analysis is based on a small number of stars and the range of derived companion masses and orbital periods is unfortunately not well constrained. 
This makes it difficult to compare the observed results with the mass distribution of the study of \citet{duq91}, for example. However, it is interesting to note 
that among the 7 targets, two seem to have very low mass companions. Whether this fits into the mass distribution of \citet{duq91} is to be confirmed once more 
systems are studied. As we said before, for every star in our sample we cannot affirm absolutely the mass and the period but only a possible range of solutions. 
Much better constraints will come from {\itshape Gaia} astrometry, that will likely allow us to find the exact mass and orbital parameters for the systems studied here. 
For example, the duration  of the {\itshape Gaia} mission will be 5 years and so it will be possible to resolve the binary system CD-436810, since this system has 
a period ($\sim6.5$ years) likely not significantly exceeding the Gaia mission duration. The fraction of astrometric binaries will also dramatically increase when 
the ({\itshape Gaia}) catalogue will be finally published. \\
In the future, thanks to the ${\mu}as$-level precision of Gaia data, it will also be possible to apply the same methods used in the present paper 
to study extrasolar planets detected both astrometrically and with Doppler measurements.

\acknowledgments
This work was supported by the Gaia Research for European Astronomy Training (GREAT-ITN) Marie Curie network, funded through the European Union Seventh Framework Programme ([FP7/2007-2013]) under grant agreement number 264895). 
This work was supported in part by the European Research Council/European Community under the FP7 through Starting Grant agreement number 239953.
NCS also acknowledges the support in the form of a Investigador FCT contract funded by Funda\c{c}\~ao para a Ci\^encia e a Tecnologia (FCT) /MCTES (Portugal) and POPH/FSE (EC).

\clearpage


%

%

%

%


\clearpage

\begin{sidewaystable*}
\begin{center}
\caption{The sample of metal-poor spectroscopic binaries included in this study.}
\label{table:1}
\begin{tabular}{c c c c c c c}
\tableline \tableline
Star & Comment & Fe/H (dex) & $M_{1}$ ($M_{\sun}$) & source of RVs & source of $[Fe/H]$ & source of $M_{1}$\\
\tableline
CD-436810 & SB1 & $-0.44$ & 0.91 & HARPS Spectrograph & \citet{adi12} & \citet{sou11}\\
HD16784 & SB1 & $-0.65$ & 0.83 & HARPS Spectrograph & \citet{sou11} & \citet{sou11}\\
G27-44 & SB1 & $-0.78$ & 0.85 & HIRES Spectrograph & \citet{soz09} & \citet{soz09}\\
G63-5 & SB1  & $-0.62$ & 0.83 & HIRES Spectrograph & \citet{soz09} & \citet{soz09}\\
G135-46 & SB1 & $-0.62$ & 0.84 & HIRES Spectrograph & \citet{soz09} & \citet{soz09}\\
G237-84 & SB1 & $-0.66$ & 0.79 & HIRES Spectrograph & \citet{soz09} & \citet{soz09}\\
HD7424 & SB1 & $-0.76$ & 0.82 & HIRES Spectrograph & \citet{soz09} & \citet{soz09}\\
HD192718 & SB1 & $-0.63$ & 0.88 & HIRES Spectrograph & \citet{soz09} & \citet{soz09}\\
\tableline
\end{tabular}
\end{center}
\end{sidewaystable*}

\clearpage


\begin{table}
\caption{List of the sample with available radial-velocity RMS from the CfA DS.}             
\label{table:RMS}      
\centering          
\begin{tabular}{c cc ccc}     
\tableline\tableline       
Star & RMS (km s$^{-1}$) \\
\tableline             
G27-44 & 0.39 \\
G63-5 & 0.36 \\
G237-84 & 0.73\\
HD7424 & 0.51 \\
HD192718 & 0.92 \\
\tableline                  
\end{tabular}
\end{table}


\clearpage


\begin{table*}[p]
\caption{List of the sample analyzed and results}             
\label{table:3}      
\centering          
\begin{tabular}{c cc ccc}     
\tableline\tableline       
Star & \multicolumn{2}{c}{Mass ($M_{\sun}$)} & \multicolumn{2}{c}{Period (years)} & Minimum Period (years) \\
\tableline  
 & 1 $\sigma$ & 2 $\sigma$ & 1 $\sigma$ & 2 $\sigma$ &  \\ 
\tableline
HD16784 & 0.53 & 0.74 & 2.2 & 3.2 & 1\\             
G27-44 & 0.05 & 0.25  & 61 & 91 & 8 \\
G63-5 & 0.07 & 0.28  & 65 & 97 & 8 \\
G135-46 & 0.54 & 0.73 & 39 & 63 & 3 \\
G237-84 & 0.1 & 0.35 & 60 & 90 & 7 \\
HD7424 & 0.43 & 0.75 & 19 & 35  & 1 \\
HD192718 & 0.55 & 0.75 & 53 & 83 & 5 \\
\tableline                  
\end{tabular}
\end{table*}
%

\clearpage

\begin{table*}[p]
\caption{Elements of the fitted orbit for the binary G135-46 and CD-436810}
\label{table:4}
\centering
\begin{tabular}{c c c c}
\tableline\tableline
 & G135-46 & CD-436810 & \\
\tableline
P & 10378.14963 $\pm$ 343.58878 & 1702.706837 $\pm$ 983.189055 & days \\
e & 0.24669 $\pm$ 0.03726 & 0.244466 $\pm$ 0.311647 &  \\
$\gamma$ & -47.1821 $\pm$ 0.1555 & 162.4862 $\pm$ 0.7259 & km s$^{-1}$ \\
$\omega$ & 107.73555 $\pm$ 8.28467 & -138.691965 $\pm$ 32.953857 & deg\\
K & 1970.701 $\pm$ 143.278 & 5373.413773 $\pm$ 588.369727 & m s$^{-1}$ \\
$T_{0}$ & 56969.75177 $\pm$ 182.83846 & 54307.191445 $\pm$ 750.931749 & JD\\
$\Delta$$RV_{T-CfA}$ & +0.316 $\pm$ 0.252 & & km s$^{-1}$\\
$\Delta$$RV_{H-CfA}$ & -0.3114 $\pm$ 0.223 & & km s$^{-1}$\\
\tableline
\end{tabular}\\
\end{table*}

\clearpage


\begin{table*}[p]
\caption{Summary of the results for the binary system CD-436810 based on the old and new Hipparcos reduction}             
\label{table:CD}      
\centering          
\begin{tabular}{c ccc ccc}     
\tableline\tableline       
Star & \multicolumn{2}{c}{Mass ($M_{\sun}$)} & Minimum Mass ($M_{\sun}$) & \multicolumn{2}{c}{Period (years)} & Minimum Period (years) \\
\tableline 
 & 1 $\sigma$ & 2 $\sigma$ & &  1 $\sigma$ & 2 $\sigma$ & \\
\tableline                  
CD-436810 old & 0.63 & 0.78 & 0.29 & 4.8 & 6 & 2.9 \\
CD-436810 new & 0.67 & 0.77 & 0.32 & 4.8 & 5.8 & 3.2 \\
\tableline                  
\end{tabular}
\end{table*}
%

\clearpage

\begin{sidewaystable*}
\caption{List of the binaries with the values of proper motion, $\Delta\mu,$ and $\dot{\mu}$ based on Hipparcos and Tycho-2, and the RV slopes based on 
available Doppler information.}\label{table:TOT}
\resizebox{0.8\textwidth}{!}{\begin{minipage}{\textwidth}
\begin{tabular*}{\tablewidth}{lcccccccr}             
\bf NAME & \bf $\pi$ (mas) & & \bf $\mu$ Hipp (mas yr$^{-1}$) & \bf $\mu$ Hipp new (mas yr$^{-1}$) & \bf $\mu$ Tycho-2 (mas yr$^{-1}$) & \bf $\Delta\mu$ (mas yr$^{-1}$) & \bf $\dot{\mu}$ (mas yr$^{-2}$) & slope (\bf m s$^{-1}$ yr$^{-1}$)  \\
CD-436810 & 9.27 & $\alpha$ & -27.20 $\pm$ 1.07 & -25.22 $\pm$ 0.94 & -18.3 $\pm$ 1.6 & 8.9 $\pm$ 1.9 & 11.29 $\pm$ 2.16 & -4885.42 $\pm$ 46.15 \\
        & & $\delta$ & -235.12 $\pm$ 1.17 & -234.14 $\pm$ 0.89 & -238.0 $\pm$ 1.2 & 2.88 $\pm$ 1.7 & &\\
HD7424 & 8.64 & $\alpha$ & 198.43 $\pm$ 2.06 & 197.59 $\pm$ 1.87 & 201.3 $\pm$ 1.6 & 2.87 $\pm$ 2.61 & & -417.02 $\pm$ 4.66\\
        & & $\delta$ &-108.32 $\pm$ 1.82 & -110.10 $\pm$ 1.49 & -114.8 $\pm$ 1.5 & 6.48 $\pm$ 2.36 & &\\
G237-84 & 29.07 & $\alpha$ &-294.31 $\pm$ 0.63 & -294.40 $\pm$ 0.57 & -300.1 $\pm$ 1.1 & 5.79 $\pm$ 1.27 & & 23.58 $\pm$ 1.72\\
        & & $\delta$ & 244.42 $\pm$ 0.66 & 244.96 $\pm$ 0.51 & 246.5 $\pm$ 1.1 & 2.08 $\pm$ 1.28 & &\\
G63-5 & 16.36 & $\alpha$ & -520.57 $\pm$ 1.13 & -520.03 $\pm$ 1.03 & -521.5 $\pm$ 1.1 & 0.93 $\pm$ 1.58 & & 12.37 $\pm$ 1.07\\
        & & $\delta$ & 267.23 $\pm$ 0.81 & 267.36 $\pm$ 0.68 & 269.4 $\pm$ 1.1 & 2.17 $\pm$ 1.36 & &\\
G135-46 & 12.96 & $\alpha$ & -334.28 $\pm$ 1.16 & -333.73 $\pm$ 1.14 & -334.0 $\pm$ 0.9 & 0.28 $\pm$ 1.47 & & 198.65 $\pm$ 2.14\\
        & & $\delta$ & -73.27 $\pm$ 1.02 & -73.97 $\pm$ 1.10 & -71.1 $\pm$ 0.9 & 2.17 $\pm$ 1.36 & &\\
HD192718 & 17.28 & $\alpha$ & 313.17 $\pm$ 1.20 & 314.39 $\pm$ 0.90 & 311.9 $\pm$ 1.5 & 1.27 $\pm$ 1.92 & & -144.65 $\pm$ 1.40\\
        & & $\delta$ & -129.31 $\pm$ 0.78 & -129.51 $\pm$ 0.61 & -133.1 $\pm$ 1.5 & 3.79 $\pm$ 1.69 & &\\
G27-44 & 23.66 & $\alpha$ & 150.64 $\pm$ 1.11 & 151.6 $\pm$ 0.69 & 150.60 $\pm$ 1.0 & 0.04 $\pm$ 1.49 & & -11.20 $\pm$ 0.78\\
        & & $\delta$ & 331.61 $\pm$ 0.75 & 331.35 $\pm$ 0.55 & 332.4 $\pm$ 1.1 & 0.79 $\pm$ 1.33 & &\\
HD16784 & 15.67 & $\alpha$ & 569.90 $\pm$ 0.95 & 569.56 $\pm$ 0.98 & 570.0 $\pm$ 1.2 & 0.1 $\pm$ 1.5 & & 10525.488 $\pm$ 6169.67 \\
        & & $\delta$ & 75.41 $\pm$ 0.90 & 75.63 $\pm$ 0.86 & 75.1 $\pm$ 1.2 & 0.31 $\pm$ 1.5 & \\
\end{tabular*}
\end{minipage}}
\end{sidewaystable*}


\clearpage

\begin{center}
\begin{longtable}{c c c}
\caption[RV measurements]{HARPS, CfA and TRES radial velocity measurements.} \label{table:m} \\

\tableline \multicolumn{1}{c}{\textbf{BJD - 2400000}} & \multicolumn{1}{c}{\textbf{Radial Velocity (km s$^{-1}$)}} & \multicolumn{1}{c}{\textbf{$\sigma$ (km s$^{-1}$)}} \\ \tableline 
\endfirsthead

\multicolumn{3}{c}%
{{\bfseries \tablename\ \thetable{} -- continued from previous page}} \\
\tableline \multicolumn{1}{c}{\textbf{BJD - 2400000}} &
\multicolumn{1}{c}{\textbf{Radial Velocity (km s$^{-1}$)}} &
\multicolumn{1}{c}{\textbf{$\sigma$ (km s$^{-1}$)}} \\ \tableline 
\endhead

\tableline \multicolumn{3}{r}{{Continued on next page}} \\ \tableline
\endfoot

\tableline \tableline
\endlastfoot

& CD$-$436810 (HARPS) & \\
& & \\
53016.845925 & 166.40259 & 0.00314\\	
53052.734014 & 166.61746 & 0.00304\\	
53056.775907 & 166.63784 & 0.00240\\	
53064.792500 & 166.68113 & 0.00273\\	
53490.601963 & 165.23422 & 0.00248\\	
53491.671209 & 165.22884 & 0.00253\\	
53492.558667 & 165.22192 & 0.00165\\	
53573.459353 & 164.42581 & 0.00357\\	
54173.743325 & 156.14358 & 0.00221\\
& & \\
& & \\
& HD16784 (HARPS) & \\
& & \\
52944.692538 & 30.56387 & 0.00297\\
53206.921719 & 36.76851 & 0.00092\\	
53216.891256 & 38.99693 & 0.00093\\
& &\\
& &\\
& G27-44 (CfA DS) &\\
& & \\
45934.6884 & -34.23  & 0.31\\
45958.5819 & -33.92 & 0.44\\
46282.7861 & -33.71 &  0.29\\
46339.7367 & -33.26 &  0.25\\
46659.6928 & -33.57 &  0.54\\
47007.8692 & -33.45 &  0.49\\
47375.8000 & -34.13 &  0.58\\
47694.9712 & -33.61 &  0.26\\
48082.8128 & -33.75 &  0.49\\
48143.6573 & -33.60 &  0.51\\
48435.8309 & -32.90 &  0.68\\
48900.6157 & -34.20 & 0.42\\
49908.8241 & -34.18 & 0.45\\
50739.6310 & -34.29 & 0.36\\
50742.6880 & -33.43 & 0.22\\
50755.6097 & -33.86 & 0.28\\
& & \\
& & \\
& G63-5 (CfA DS) & \\
& &\\
45037.9649 &  5.88 & 0.17\\
45717.8215 &  5.64 & 0.24\\
45860.6643 &  5.08 & 0.18\\
46509.8467 &  5.92 & 0.27\\
46843.8410 &  5.84 & 0.19\\
47201.9465 &  5.82 & 0.31\\
47549.9644 &  5.92 & 0.20\\
47963.8947 &  5.04 & 0.37\\
48371.6681 &  5.91 & 0.51\\
& & \\
& & \\
& G237-84 (CfA DS) & \\
& &\\
45772.7879  &  9.78 &  0.18\\
45833.8371  &  8.09 &  0.61\\
45886.6972  &  7.08 &  0.21\\
46107.9933  &  9.11 &  0.17\\
46577.6817  &  9.13 &  0.29\\
46953.6721  & 10.48 & 0.31\\
46986.6346  &  9.44 & 0.25\\
47134.9914  & 10.13 &   0.37\\
47159.0571 &   8.99  &  0.23\\
47556.0699  &  8.81  & 0.22\\
47575.0687  &  8.68 &   0.21\\
47602.7910  &  8.64 &  0.17\\
47604.8590  &  8.81  &  0.22\\
47629.8029  &  9.04 &  0.23\\
48288.9797  &  9.24 &  0.23\\
48605.0531  &  9.09 & 0.22\\
49056.9222  &  9.05 & 0.38\\
54574.7592  &  9.88 &  0.43\\
54608.7376  &  9.74 & 0.40\\
54844.9614  & 10.67 &  0.25\\
54901.8969  &  9.88  & 0.50\\
54929.7681  & 10.11  &   0.40\\
55198.9976  &  9.51  & 0.30\\
55284.8971  &  9.82  & 0.20\\
& & \\
& G237-84 (TRES) & \\
& &\\
55311.8115  &  9.69 &  0.10\\
55340.6311  &  9.72 &  0.10\\
55584.0480  &  9.74 &  0.10\\
55666.8658  &  9.79 & 0.10\\
55668.7458  &   9.76 & 0.10\\
& & \\
& & \\
& HD7424 (CfA DS) & \\
& &\\
45722.6160 &  83.94  &  0.44\\
45960.7864 &  85.79  &   0.97\\
46673.7901 &  84.93  &   0.78\\
47018.8042 &  84.78  &   0.77\\
47431.9151 &  85.65  &  0.38\\
47780.7328 &  84.57  &   0.68\\
48856.8328 &  84.80  &   0.53\\
50711.8584 &  84.65  &   0.38\\
50742.7854 &  85.04  &   0.27\\
50744.7392 &  84.83  &   0.22\\
& & \\
& HD7424 (TRES) & \\
& &\\
55242.5827 &  84.56  &   0.10\\
& & \\
& & \\
& HD192718 (CfA DS) & \\
& &\\
48220.4871 & -111.18 &  0.69\\
48405.8449 & -110.87 &   0.67\\
48415.8438 & -110.87 &   0.73\\
48433.7621 & -110.96 &   0.84\\
51395.7013 & -110.68 &    0.72\\
53192.9203 & -112.11 &    0.34\\
& & \\
& HD192718 (TRES) & \\
& &\\
56087.9710 & -112.73 &    0.10\\
56134.8973 & -112.71 &   0.10\\
56260.5529 & -112.81 &   0.10\\
& & \\
& & \\
& G135-46 (CfA DS) & \\
& &\\
46928.8399 & -48.63  &  0.66\\
47174.9539 & -49.67  &  0.64\\
47187.9622 & -48.32  &   0.91\\
47225.0309 & -49.35  &   0.60\\
47634.8100 & -48.14  &  0.47\\
48374.7499 & -49.87  &  0.60\\
49115.8432 & -48.81  &   0.52\\
49470.8184 & -48.88  &   0.46\\
49478.7951 & -47.86  &  0.46\\
50093.9574 & -47.86  &   0.46\\
53190.6938 & -46.05  &  0.27\\
53216.6161 & -46.05  &  0.68\\
53251.5128 & -45.66  &   0.47\\
53343.9478 & -45.91  &  0.59\\
53399.9381 & -45.55  &  0.52\\
53434.8018 & -46.58  &  0.39\\
53480.7413 & -44.96  &  0.42\\
53513.8344 & -45.91  &   0.39\\
53812.9219 & -45.49  &  0.54\\
53871.8073 & -45.73  &   0.24\\
54134.0428 & -45.05  &   0.34\\
54160.0345 & -44.93  &   0.36\\
54193.8909 & -46.15  &  0.32\\
54219.8339 & -45.52  &  0.31\\
54485.0612 & -45.06  &  0.46\\
54548.9753 & -45.40  &   0.43\\
54575.8548 & -45.85  &  0.42\\
54575.9277 & -45.65  &  0.48\\
54605.7664 & -45.78  &   0.40\\
54632.7114 & -45.37  &    0.40\\
54843.0356 & -45.08  &   0.38\\
54929.8908 & -46.01  &  0.61\\
54930.9282 & -45.00  &   0.62\\
54961.8243 & -45.76  &   0.36\\
55251.9779 & -44.62  &  0.60\\
55284.9205 & -45.36  & 0.30\\
55731.7849 & -45.42  &   0.20\\
& & \\
& G135-46 (TRES) & \\
& &\\
55200.0496 & -45.19  &  0.10\\
55308.9105 & -45.09  &  0.10\\
55344.7601 & -45.29  &  0.10\\
55577.0259 & -45.36  &  0.10\\
55665.8452 & -45.40  &  0.10\\
55692.7503 & -45.35  &  0.10\\
55960.0303 & -45.84  & 0.10\\
55990.9158 & -45.88  &  0.10\\
56024.8898 & -45.91  & 0.10\\
56047.8309 & -45.96  & 0.10\\
56089.6405 & -45.95  &   0.10\\
56140.6684 & -46.08  &   0.10\\
56288.0453 & -46.21  & 0.10\\
56309.0569 & -46.42  &  0.10\\
56348.0213 & -46.35  &  0.10\\
56377.8662 & -46.51  &   0.10\\
& & \\
\end{longtable}
\end{center}

\end{document}